\documentclass[12pt]{iopart}
\usepackage{iopams}

\pdfoutput=1
\usepackage[pdftex]{graphicx}
\usepackage[pdftex]{color}

\usepackage[colorlinks,bookmarks]{hyperref}
\definecolor{linkblue}{rgb}{0,0,0.8}
\definecolor{linkgreen}{rgb}{0,0.5,0}
\hypersetup{linkcolor=linkblue, citecolor=linkgreen}

\begin{document}

\title[Observational constraints on inhomogeneous models without dark energy]{Observational constraints on inhomogeneous cosmological models without dark energy}

\author{Valerio Marra$^1$ and Alessio Notari$^2$}

\address{$^1$Department of Physics, PL 35 (YFL), FI-40014 University of Jyv\"askyl\"a, Finland, and
Helsinki Institute of Physics, PL 64, FI-00014 University of Helsinki, Finland}

\address{$^2$Departament de F\'isica Fonamental i 
Institut de Ci\`encies del Cosmos, Universitat de Barcelona, 
Mart\'i i Franqu\`es 1, 08028 Barcelona, Spain}

\ead{valerio.marra@me.com {\rm and} notari@ffn.ub.es}

\begin{abstract}
It has been proposed that the observed dark energy can be explained away by the effect of large-scale nonlinear inhomogeneities. In the present paper we discuss how observations constrain cosmological models featuring large voids. We start by considering Copernican models, in which the observer is not occupying a special position and homogeneity is preserved on a very large scale. We show how these models, at least in their {\em current} realizations, are constrained to give small, but perhaps not negligible in certain contexts, corrections to the cosmological observables.
We then examine non-Copernican models, in which the observer is close to the center of a very large void. These models can give large corrections to the observables which mimic an accelerated FLRW model. We carefully discuss the main observables and tests able to exclude them.
\end{abstract}



\section{Introduction}

A striking feature of the late universe is the abundance of large voids seen both in galaxy redshift surveys \cite{Hoyle:2003hc, Frith:2003tb, Tikhonov:2007di, Labini:2009nx,sylosFOCUS} and large-scale simulations \cite{Springel:2005nw}.
Voids dominate the total volume of the universe while matter is mostly distributed into a filamentary structure dubbed as the cosmic web.
Photons, therefore, travel mostly through voids and it has been asked by various authors if large-scale nonlinear inhomogeneities can sizeably alter the observables as compared to the corresponding homogeneous and isotropic Friedmann-Lema\^{i}tre-Robertson-Walker (FLRW) model.
The study of the differences between the ``optical universe'' and the homogeneous counterpart started perhaps with Zel'Dovich \cite{zel64} in the 1964 and continued in the following years by, for example, Refs.~\cite{bertotti66, dashe66, gunn67, Kantowski:1969, DyerRoeder72, Weinberg:1976jq, Ellis:1998ha}.

After 1998, when the dark energy became observationally necessary \cite{Riess:1998cb, Perlmutter:1998np}, it was then asked (see, for example, \cite{Celerier:1999hp,Tomita:1999qn,Rasanen:2003fy, Kolb:2004am,Buchert:2007ik} and references therein) if the corrections could be large enough so to have a paradigm shift in which the dark source is no longer needed.
The large-scale structures became indeed nonlinear recently, exactly when a primary dark energy is supposed to start dominating the energy content of the universe and cause acceleration.
In this paper we will review some of the contributions to this topic and how they are constrained by experimental data. The common thread will be the modeling of voids.

The effect of large-scale inhomogeneities on cosmological observables has been studied in the literature following two {\em non-exclusive} approaches.
With the first it is assumed that one can compute observational quantities as in an effective homogeneous and isotropic background metric, whose evolution equations however are modified, as compared to the usual Friedmann equations,
by the presence of large-scale structures due to the nonlinear nature of gravity;
this topic will be covered in detail by other reviews of this Special Issue~\cite{kolbFOCUS, buchertFOCUS, ellisFOCUS, chrisFOCUS, wiltshireFOCUS, rasanenFOCUS, phasespaceFOCUS,ClarksonFOCUS}.
The second approach, on the other hand, focuses on the observational properties of the universe: even in the case in which one ignores the nonlinear effects of gravity on the overall background evolution, there are still potential differences due to the fact that photons in a inhomogeneous universe propagate differently than in an FLRW model.
The former effect has been called sometimes ``strong'' backreaction and the latter ``weak'' backreaction (see Ref.~\cite{Kolb:2009rp} for more precise definitions).
As will be clear later on, the models discussed in this paper address the weak backreaction issue, as strong backreaction is by construction small in these examples.

This paper is organized as follows.
In Section \ref{coper} we start by considering Copernican models, in which the observer is not occupying a special position and homogeneity is preserved on a very large scale.
We then examine in Section \ref{noncoper} non-Copernican models in which the observer is close to the center of a very large void.
We give our conclusions and discuss future prospects in Section \ref{conclusions}.
In \ref{LTB} we briefly introduce the spherically-symmetric LTB metric and in \ref{curvatureLTB} we argue that strong backreaction effects are small for realistic exact Swiss-cheese models.

\section{Copernican Models} \label{coper}

Within Copernican models the observer is not occupying a special position and the corrections to the observables come from the cumulative effect of many inhomogeneities.
In this Section we will consider two particular configurations, namely, the so-called ``swiss-cheese'' and ``meatball'' models.

\subsection{Swiss-Cheese Models}

The first swiss-cheese models go back to Einstein and Strauss and consist of Schwarzschild holes embedded into an Einstein-de Sitter (EdS) background \cite{einstein45}, and the idea was to hide matter from the observer by confining it into the Schwarzschild masses at the center of the swiss-cheese holes \cite{Kantowski:1969}.
This idea will be developed with the meatball models in the next Section, while here we will focus on the modelling of the voids by considering a swiss cheese with LTB holes.
Lema\^{i}tre-Tolman-Bondi models \cite{Lemaitre:1933gd,Tolman:1934za,Bondi:1947av} are spherically symmetric dust solutions of Einstein's equations with pure radial motion and without shell crossing (see \ref{LTB} for a brief introduction).

The solutions we are considering involve only pressureless matter. However, this should not affect the modelling of the voids at late times as pressure is negligible within them (see \cite{Bolejko:2005tk,Lasky:2010vn} for models which contain also radiation).
Moreover, the spherical approximation should not be too unrealistic as voids tend to evolve towards a spherical configuration. It can indeed be shown with the top-hat void model that the smaller axis of an underdense ellipsoid grows faster as compared to the longer ones, with the consequence that voids become increasingly spherical as they evolve \cite{Sheth:2003py}.
The exact opposite happens to overdense ellipsoids which tend to form filamentary and pancake-like structures.
On the other hand, simple spherical voids neglect the possible presence of substructure and the interactions between the underdensities.

\begin{figure}
\begin{center}
\includegraphics[height= 6 cm]{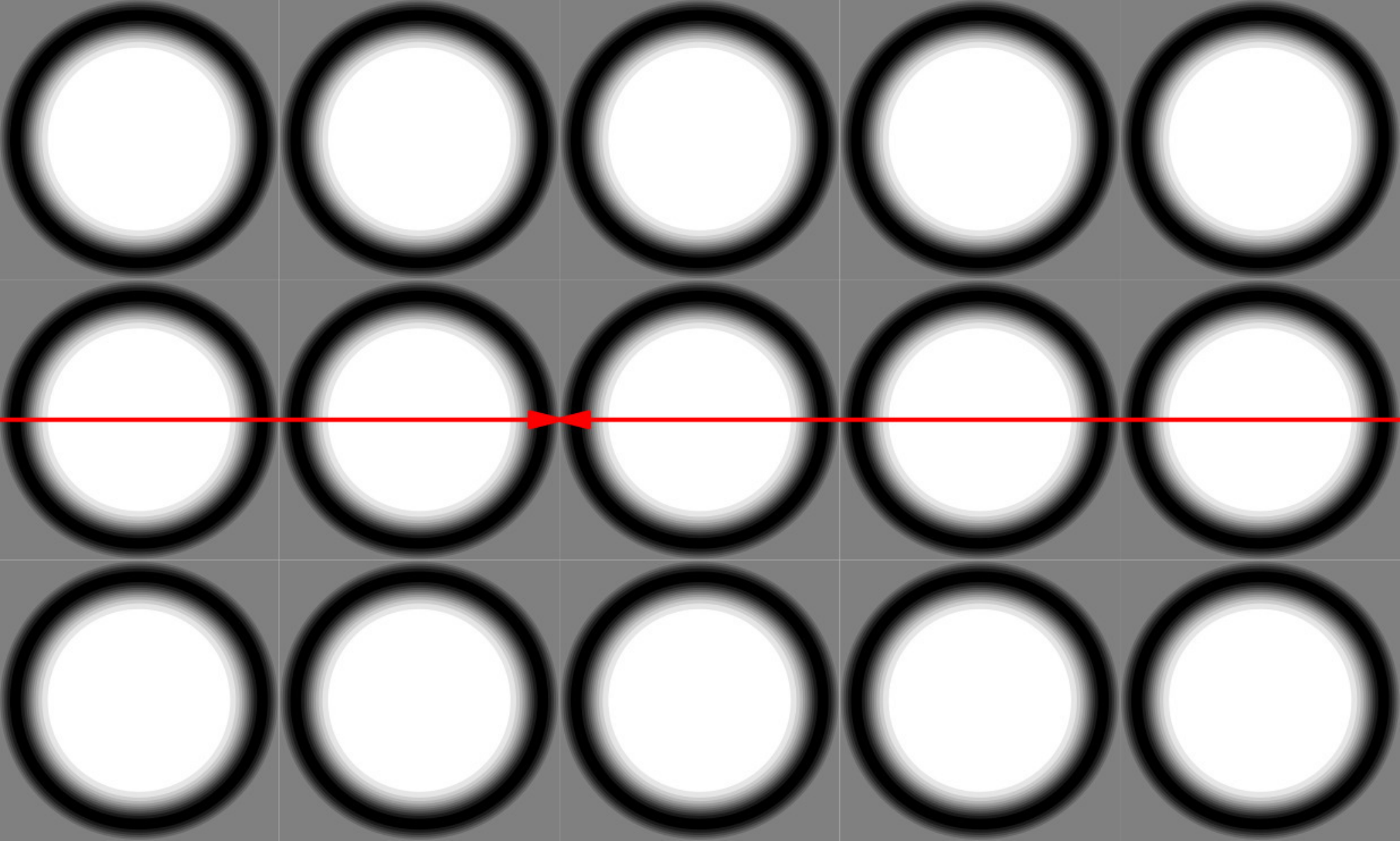}
\caption{Sketch of the Swiss-cheese model of Ref.~\cite{Marra:2007pm, Marra:2007gc}. The shading mimics 
the density profile: darker shading implies larger densities. The uniform 
gray is the EdS cheese. The photons pass through the holes as shown by the 
arrows and are revealed by the observer placed in the cheese.}
\label{schizzo}
\end{center}
\end{figure}

The general setup of an LTB patch is of an inner underdensity (``hole'') surrounded by a compensating overdensity (``crust''), necessary to match the EdS background (``cheese''): see, for example, the sketch of Fig.~\ref{schizzo}.
A physical picture is that, given an EdS sphere, all the matter in the inner region is pushed to the border of the sphere while the quantity of matter
inside the sphere does not change (note that this is valid only for holes matched at a finite radius).
More precisely, in order to match exactly the two metrics the average density of the inhomogeneous region has to be equal to the density of the external FLRW metric.\footnote{A subtlety is that the average density has to be defined with a volume element which does not include the curvature term (see the discussion about the effective mass function $F(r)$ in \ref{LTB} and also Ref.~\cite{Bondi:1947av, Sussman:2011na} ) .}
With the density chosen in this way, an observer outside the hole will not feel its presence as far as \textit{local} physics is concerned (this does not apply to global quantities, such as the distance--redshift relation for example). So the cheese is evolving as an EdS universe while the holes evolve differently.
In this way we can imagine putting in the cheese as many holes as we want, even with different sizes and density profiles, and still have an exact solution of Einstein's equations (as long as there is no superposition among the holes and the correct matching is achieved).
This property is due to spherical symmetry and it is qualitatively analogous to Gauss' theorem in electrodynamics or to Birkhoff's theorem for black holes.
While this allows to solve the equations exactly, it eliminates on the other hand any important ``strong'' backreaction~\cite{Kolb:2009rp} effect in the FLRW region, which keeps evolving {\it exactly} as in a pure FLRW model no matter how the hole is modeled.
For more details see \ref{curvatureLTB} where we argue that strong backreaction effects are small for realistic exact Swiss-cheese models  (see also~\cite{Sussman:2011na} for a discussion about strong backreaction in LTB models which converge asymptotically to FLRW).
For this reason it may be of crucial importance to study metrics which do not have the property of being exactly glued to FLRW, while still having homogeneity and isotropy on large scales. So far, to our knowledge, all the models in the literature are based on LTB or its generalizations, and thus still preserve this property, even in absence of spherical symmetry (see, for example, Refs.~\cite{Ishak:2007rp, Bolejko:2010wc, Bolejko:2010eb} for models based on the Szekeres metric; see Ref.~\cite{Rasanen:2009uw} for models which only assume statistical homogeneity and isotropy; see Ref.~\cite{bolejkoFOCUS} of this Special Issue for an extensive discussion of inhomogeneous models relevant for cosmology).

LTB swiss-cheese models have been studied in the literature with different configurations and sizes, analytically and numerically. In order to interpret cosmological observations the key quantities to be computed for a given source are redshift and distance, as for instance in any supernova analysis. Inhomogeneities affect both observables through redshift and lensing effects, which we now discuss.

\subsubsection{Redshift Effects.} \label{redef}

When a photon crosses a hole, it gains an additional redshift $\delta z$ with respect to the usual FLRW redshift.
It is, however, natural to expect a compensation on the redshift effects between the ingoing and outgoing geodesic paths due to the spherical symmetry.
Moreover, because the LTB metric is matched to the background metric, it can be shown that there is a compensation already on the scale of half a hole.
The effect on a photon has been studied exactly in the full nonlinear LTB metric~\cite{Marra:2007pm,Biswas:2006ub,Biswas:2007gi,Brouzakis:2008uw} both numerically and using analytical approximations.
It is, however, possible to capture the qualitative meaning of the various leading effects by means of a perturbative approach which should be valid if the size of the hole is much smaller than the FLRW horizon and could perhaps give a better idea to the reader of the physics behind the exact  results.
Of course this discussion does not substitute the exact nonlinear one, which is anyway necessary when the LTB patch extends to sizes comparable to the Hubble horizon.

The correction to the redshift can be understood via the well-known approximate expression~\cite{Sachs:1967} for the redshift (or equivalently for the temperature of photons if they are blackbody distributed) in a given direction $e_i$, which is expressed here in the Newtonian gauge and is valid for small velocities and potentials:
\begin{equation} \label{dez}
{\delta z \over 1+z}= \frac{\delta T}{T} \simeq \Phi_E-\Phi_O- v^i_{E} \, e_i+v^i_{O} \, e_i - 2  \int_O^E d\tau \, \frac{\partial \Phi}{ \partial \tau} \,.
\end{equation}
In Eq.~(\ref{dez}) the first two terms give the so-called Sachs-Wolfe effect, due to the difference in gravitational potentials $\Phi$ at Emitter and Observer, the third and fourth are the Doppler effects due to the velocity $v^i$ of the Emitter and the Observer and the last term is the Integrated Sachs Wolfe (ISW) effect, due to the change in (conformal) time $\tau$ of the potential along the line of sight. Each of these terms can be explicitly computed in the LTB metric in the ``small-$u$" approximation, which has been extensively studied in~\cite{Biswas:2006ub,Biswas:2007gi}.

As long as the proper radius of the hole $l_{0}$ is much smaller than the size of the FLRW horizon  $l_{\rm hor}$, we can easily understand which are the dominant terms.
If the observer or the source occupy a generic position, the dominant terms are the velocities, which can be shown to be of order $l_{0}/l_{\rm hor}$.
If we put the source at the boundary and the observer at the centre (or viceversa), the velocities vanish, because of the exact matching to FLRW, and there are only two nonzero terms: $\Phi_{O (E)}$ and the ISW effect, of which the dominant one is $\Phi_{O (E)}$ which can be shown to be of order $(l_{0}/l_{\rm hor})^2$.
Finally, if we put {\it both} the observer {\it and} the source on the boundary, the only non-vanishing contribution is the ISW, which is further suppressed and goes as $(l_{0}/l_{\rm hor})^3$.
This last effect is also known as the Rees-Sciama effect~\cite{Rees:1968} and comes from the fact that the hole and its structures  evolve while the
photon is passing, and it has been confirmed analytically in LTB models by Ref.~\cite{Biswas:2007gi, Brouzakis:2008uw}, in which more details can be found.

The suppression of redshift effects due to the matching of the LTB metric to the EdS background can be also qualitatively understood by noting that $dz/dr=H \propto \rho = \rho_{\scriptscriptstyle EdS} + \delta \rho$.
Because the density profile satisfies $\langle \delta \rho \rangle=0$, in its journey from the center to the border of the hole (or viceversa) the photon will see a $\langle H\rangle \sim H_{\scriptscriptstyle EdS}$ and there will be compensation for $z'$.
This can also be seen by line averaging the longitudinal expansion rate $H_{L}$ (see \ref{LTB} for definitions) from the center to the boundary of the hole \cite{Marra:2007pm}:
\begin{equation} \label{hint}
\langle H_L \rangle \simeq
\frac{\int_{0}^{r_{0}}dr \; H_L \; Y' }
{\int_{0}^{r_{0}} dr \; Y' } = \left. \frac{\dot{Y}}{Y}\right|_{r=r_{0}}
= H_{\scriptscriptstyle EdS} \,,
\end{equation}
where $Y$ is a $r$-dependent generalization of the scale factor for LTB metrics, $r_{0}$ is the comoving radius of the LTB patch ($Y(r_{0},t)= a(t) \, r_{0}= l_{0}$) and the approximation comes from neglecting the small curvature $E$ and the time evolution of the density profile.
The latter can be qualitatively interpreted as neglecting the ISW effect and the former as neglecting the Sachs-Wolfe effect ($\Phi \propto E$, see \cite{Biswas:2007gi}).

\begin{figure}
\begin{center}
\includegraphics[width= .48\textwidth, height=4.7 cm]{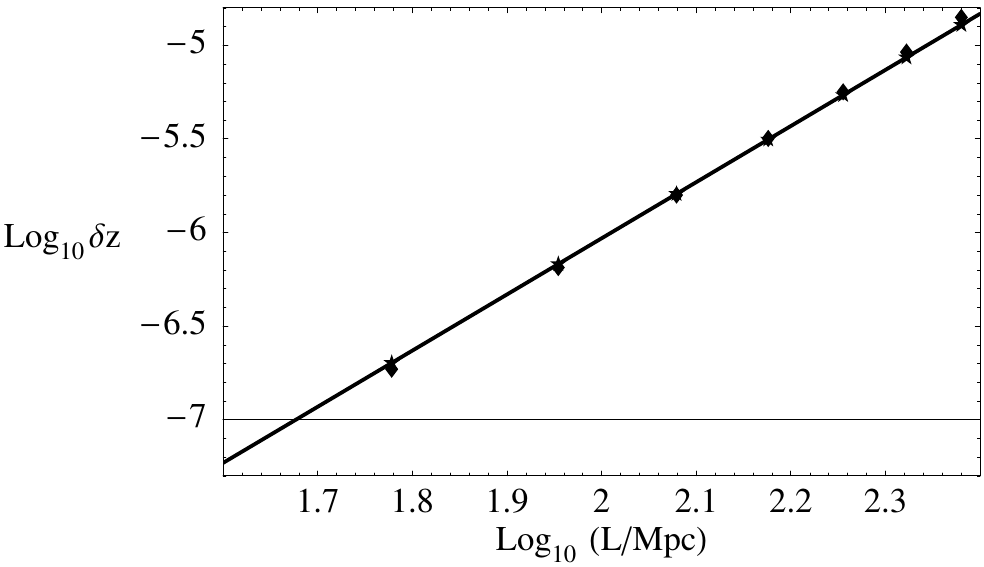}
\includegraphics[width= .48\textwidth, height=4.7 cm]{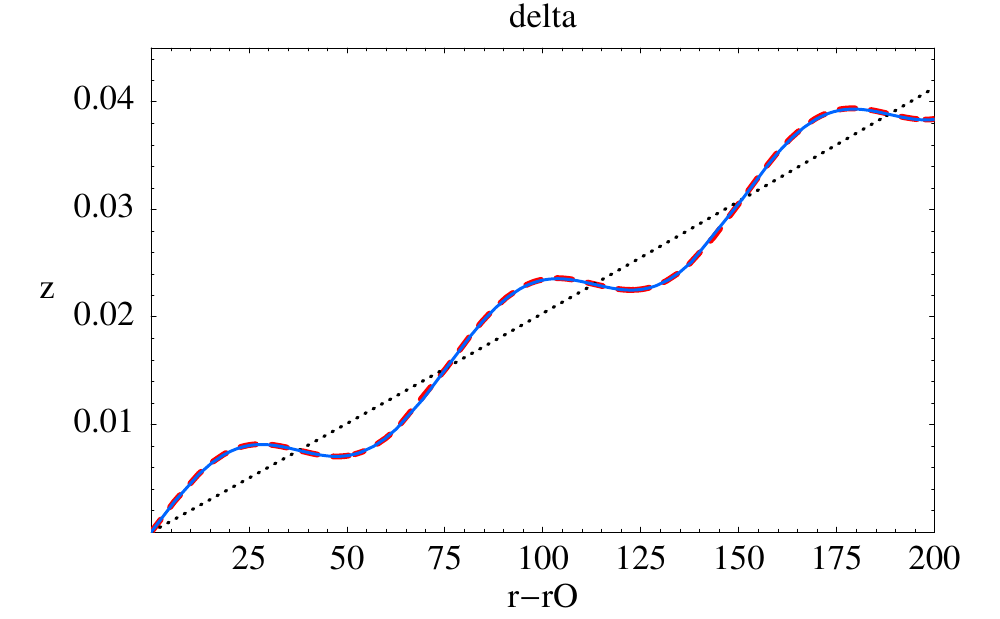}
\caption{
Left panel: net correction to the redshift ($\delta z$) for the Swiss-cheese model of Ref.~\cite{Biswas:2007gi} as a function of the size $L$ of a hole with $\delta\approx 0.2$, for an observer sitting outside the hole.
The triangles are the numerical results, while the solid line shows the predicted cubic dependence ($\delta z\propto L^3$).
Right panel: redshift along the geodesic of a photon revealed by an observer far from the center of the Onion model of Ref.~\cite{Biswas:2006ub}.
The solid line is the numerical solution, the dashed line is the analytical approximation and the dotted line is the FLRW result. 
}
\label{onion}
\end{center}
\end{figure}

In conclusion, redshift effects are large, and actually the dominant ones if one considers sources or observers within the holes, and especially at low redshift.
This has been discussed in detail for the Swiss-cheese model of Ref.~\cite{Biswas:2007gi} (left panel of Fig.~\ref{onion}), and also for the so-called Onion model of Ref.~\cite{Biswas:2006ub} (see also \cite{Marra:2008sy}), where the universe is described by spherical concentric shells with positive and negative contrast (right panel of Fig.~\ref{onion}).
Similar results can be seen also from Fig.~\ref{lumidi} (left panel), which is relative to the Swiss-cheese model of Refs.~\cite{Marra:2007pm, Marra:2007gc}. 
However these are mainly due a Doppler effect from peculiar velocities, and this should average out when considering many sources, except in the case of non-Copernican models, which will be addressed in Section \ref{noncoper} of this review. 
If we indeed consider light going through the Swiss-cheese holes and both observer and emitter in the cheese, then the effect from a single hole is suppressed as $\left(l_{0}/l_{\rm hor}\right)^3$. If now light goes through a series of many voids, the effect basically sums up: in the extreme case of filling the universe with LTB patches, a photon may meet $l_{\rm hor}/l_{0}$ number of holes and  this gives an overall correction of order $(l_{0}/l_{\rm hor})^2$, which would not be enough to explain away dark energy, for any realistic configuration with voids smaller than the horizon (see \cite{Valkenburg:2009iw} for an extensive numerical analysis).

\subsubsection{Lensing Effects.}

\begin{figure}
\begin{center}
\includegraphics[width= .48\textwidth, height=4.7 cm]{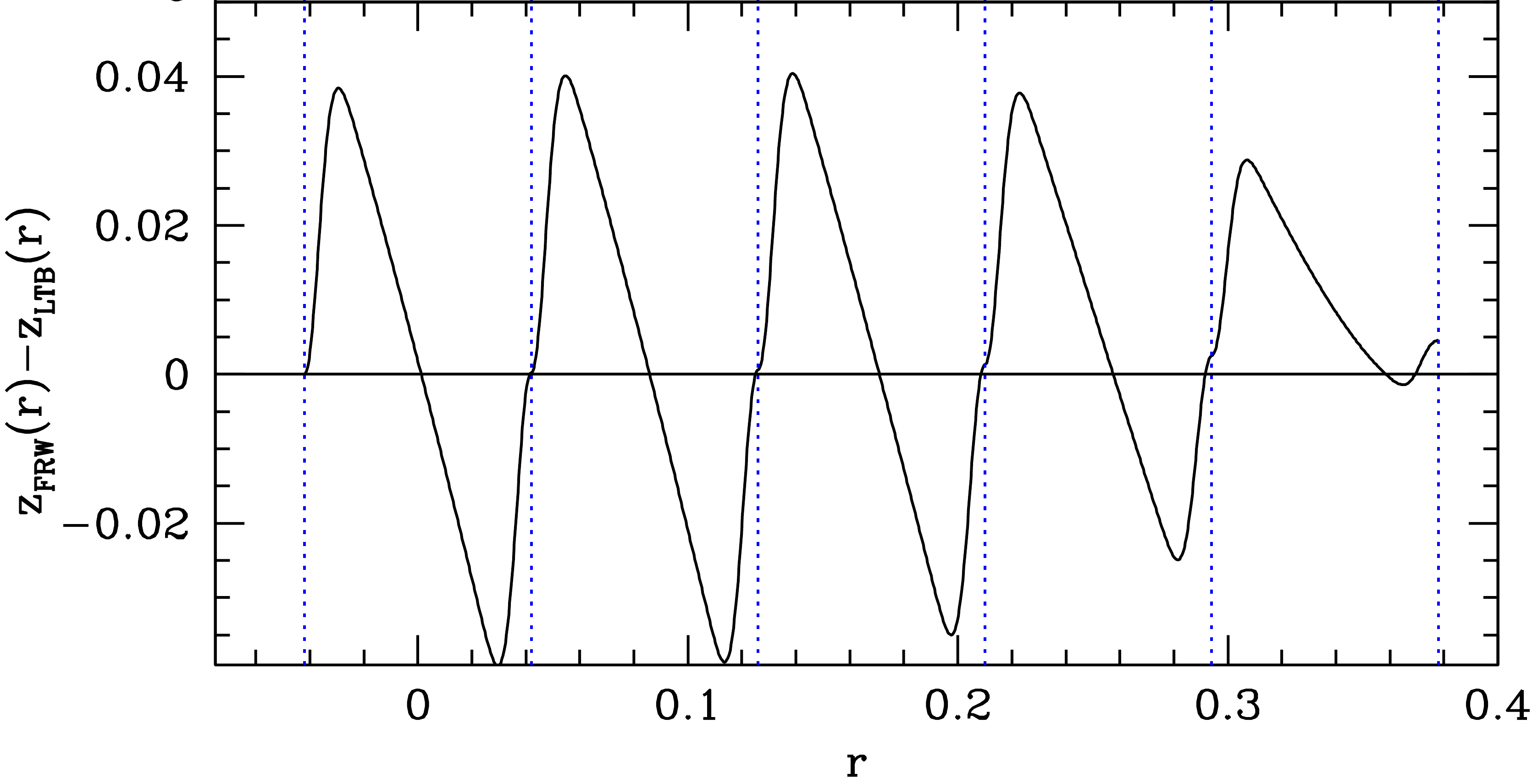}
\includegraphics[width= .48\textwidth, height=4.7 cm]{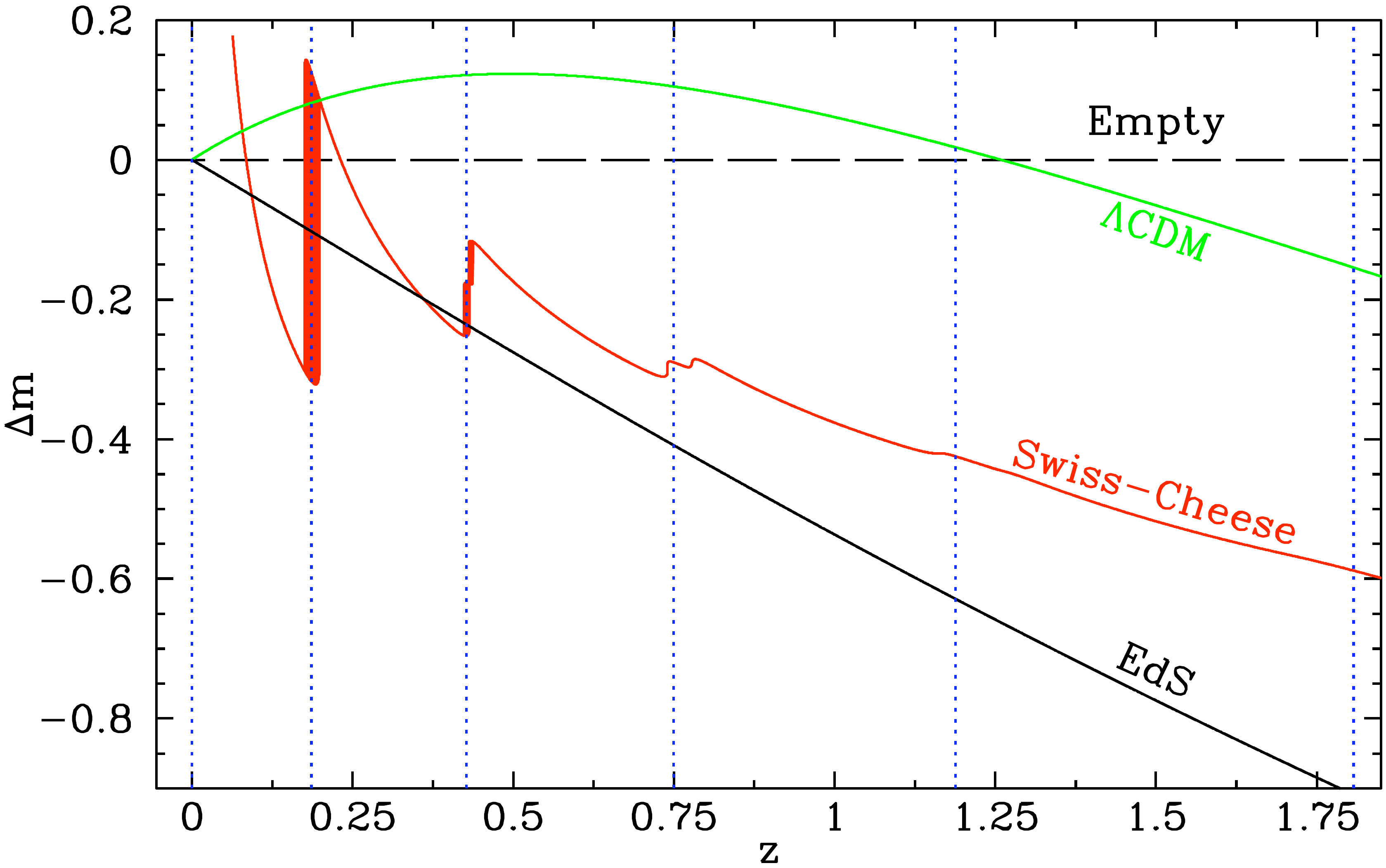}
\caption{
Left panel: change in redshift with respect to the EdS model for a photon that travels from one 
side of the five-hole chain of Fig.~\ref{schizzo} to the other where the 
observer will detect it at present time. The vertical lines mark the edges of the holes. The plots are with
respect to the coordinate radius $r$. Notice also that along the inner voids the
redshift is increasing  faster because of the higher expansion rate ($z'(r)=H(z)$).
Right panel: the luminosity distance as a function of redshift for the observer of Fig.~\ref{schizzo}, together with the EdS and $\Lambda$CDM curves. Rather than $H_0 d_L(z)$, we show the usual difference
in the distance modulus compared to the empty model.}
\label{lumidi}
\end{center}
\end{figure}

In addition to the effects on the redshift of a source, it is necessary to study the effects on its distance, which can be defined equivalently as the angular distance $d_A^2\equiv dA / d\Omega$ or the luminosity distance $d_L=(1+z)^2 \, d_A$, where $dA$ and $d\Omega$ are the area of an object and the solid angle under which it is seen by an observer, respectively.
Rather than the distances above, we will use the distance modulus which is defined as $m(z)= 5 \log_{10} d_L(z) / (10\, {\rm pc})$.
Large-scale inhomogeneities affect distance measurements because their lensing effects change the ratio $dA/d\Omega$. This has been computed by several authors in the exact LTB model by solving explicitly for the geodesics in the full LTB metric, both numerically and with analytical approximations~\cite{Marra:2007pm, Marra:2007gc,Biswas:2006ub,Biswas:2007gi,Valkenburg:2009iw,Brouzakis:2007zi, Szybka:2010ky}.
Lensing effects can be also simply understood in the weak-lensing limit of small magnifications, which we briefly introduce as follows. For more details see, for example, Ref.~\cite{Bartelmann:1999yn}.

In the weak-lensing theory the net magnification $\mu$ produced by a localised density perturbation is:
\begin{equation} \label{mag}
\mu={1 \over (1-\kappa)^{2} - |\gamma_{s}|^{2}}
\simeq (1-\kappa)^{-2}\,,
\end{equation}
where $\kappa$ is the lens convergence and we have neglected the second-order contribution of the shear $\gamma_{s}$.
The convergence is due to the local matter density (see Eq.~(\ref{k09}) below) and its physical effect is to magnify the image by increasing its size; the shear instead is due to the tidal gravitational field and its effect is to stretch the image tangentially deforming a circular shape into an elliptic one~\cite{Bartelmann:1999yn}.
The shift in the distance modulus caused by $\mu$ then becomes:
\begin{equation} 
\Delta m = -2.5 \log_{10}\mu \, \simeq \, 5 \log_{10}(1-\kappa)\,.
\label{eq:dm}
\end{equation}
The lens convergence $\kappa$ can be computed from the following integral along an unperturbed light geodesic:
\begin{equation} 
   \kappa(z_{s})=\int_{0}^{r_{s}} dr \,  \frac{r(r_{s}-r)}{r_{s}} \nabla^2 \Phi \,,
\label{k09}
\end{equation}
where $r$ labels the comoving distance and $z_{s}$ is the redshift of a light source whose comoving position is $r_{s}=r(z_{s})$.
The term $\nabla^{2} \Phi $ is the Laplacian of the Newtonian potential, given by 
\begin{equation}
 \nabla^{2} \Phi 
  = 4\pi G a^2  \; \delta \rho_{M} 
  = \frac{3}{2} \, \frac{H_0^2} {a} \,  \delta_{M} \,,
\label{eq:kappa2}
\end{equation}
where $a(t)$ is the EdS scale factor, $\delta_{M}=\delta \rho_{M}/\rho_{M}$ is the matter contrast and $\rho_{M}$ is the matter density.
From Eq.~(\ref{k09}) follows then
\begin{equation} \label{eq:kappa}
 \kappa(z_{s})=  \frac{3 }{2} H_{0}^{2}   \int_{0}^{r_{s}} dr  \,  {r (r_{s}-r) \over r_{s}} \,   {\delta_{M}(r,t(r)) \over a(t(r))} \,.
\end{equation}
The previous equations show that for a lower-than-EdS
column density the light is demagnified (e.g., empty beam $\delta_{M}=-1$),
while in the opposite case it is magnified.
The weak-lens approximation is valid when the universe can be described by a Newtonian-perturbed FLRW metric and when $\kappa \ll 1$, i.e., when the average matter contrast along the line of sight is $\delta_{M} \lesssim 1$.

As shown in Fig.~\ref{lumidi} (right panel), Refs.~\cite{Marra:2007pm, Marra:2007gc} found a large effect for the setup where an observer in the cheese looks through a chain of aligned voids (see Fig.~\ref{schizzo}). The idea was indeed that the photons we observe
have likely travelled more through (large) voids than through dense structures. The effect builds up and it becomes large at high redshift $z\gtrsim 1$.
The particular configuration chosen in~\cite{Marra:2007pm, Marra:2007gc} had holes of a radius of 350 Mpc; however, it can be shown~\cite{Valkenburg:2009iw} that the effect does not depend on the size of the voids as long as the average matter contrast along the line of sight remains approximately constant (see Eq.~(\ref{eq:kappa})).
For the setup of Fig.~\ref{schizzo}, a chain of 50 holes which are 10 times smaller will indeed yield a lightcone column density approximately equal as compared to the configuration with 5 larger holes.
For constant-time slices this is actually exactly true, since the LTB dynamics is {\em invariant} with respect to the hole radius once the density profile is properly scaled~\cite{Marra:2007pm}.

Such a result, however, can be valid only for particular directions in the sky, such as the one where all the centers of the voids are aligned, and not for a generic direction.
Indeed, since the distance is a measure of the total luminosity received and the number of photons is conserved, the average of $d_L$ and therefore $d_A$ over the full sky must be identical\footnote{This should be approximately true~\cite{Weinberg:1976jq} if strong lensing events leading to secondary images and caustics~\cite{Ellis:1998ha} do not play a significant role as far as the full-sky average is concerned~\cite{Holz:1997ic,Hilbert:2007ny,Hilbert:2007jd}.} to the unperturbed sky, at least in the case of weak lensing where no photons are lost along the path\footnote{This can be seen by the fact that it is $\langle \delta_{M} \rangle =0$ and so  Eq.~(\ref{eq:kappa}) implies $\langle \kappa \rangle =0$.}. In other words, while some areas of the sky are demagnified other areas are magnified, yielding an exact compensation for the full-sky average, showing the ``benevolent'' nature of weak lensing.  So, the only overall net effect for the full-sky average is in principle the change in the photon redshift discussed in the previous Section.
This has been confirmed by Refs.~\cite{Valkenburg:2009iw, Szybka:2010ky,Vanderveld:2008vi} (see also \cite{Brouzakis:2007zi}) where, by averaging over the lines of sight or by randomizing the positions of the holes, the luminosity distance has been found to converge to the EdS prediction.
An additional relevant result is that the Rees-Sciama effect on the CMB constrains the holes of a swiss-cheese model to have a radius smaller than about 35 Mpc~\cite{Valkenburg:2009iw}, thus making it unlikely that a photon always travelled through the central regions of the voids as in Fig.~\ref{schizzo}.

It is interesting to note that, even if the average is preserved, the lensing probability distribution function (PDF) is modified by the presence of structures. 
The Swiss-cheese model, however, is not suitable to study the statistical properties of light propagation and the basic reason is that a photon always hits the surrounding overdensity before and after passing through the underdensity. This geometrical feature imposed by the need to match the EdS metric constrains the model to have a nearly gaussian PDF~\cite{Valkenburg:2009iw,Brouzakis:2007zi,Vanderveld:2008vi}, especially if the voids, as discussed above, have to be smaller than about 35 Mpc.
In order to properly study the statistical properties of the luminosity distance in a universe dominated by voids, i.e., the lensing PDF, it is necessary therefore to focus on a model that allows photons to miss overdensities: this will be the subject of the next Section.

\subsection{Meatball Models} \label{mmo}

The meatball model (see, for example, \cite{mbTopology,Kainulainen:2009sx} and also the lattice model of \cite{Clifton:2009jw}) describes the universe as made of a collection of possibly different spherical halos\footnote{
See Ref.~\cite{Amendola:2010ub} for a reanalysis of the SNe data which includes the lensing effects of the meatball model.} and incorporates quantitatively the crucial feature that photons can travel through voids, miss the localized overdensities and experience a low matter column density.
This is due to the fact that the underdensities occupy more volume than the overdensities and causes the lensing PDF to be skewed with a mode at demagnified values.
This is of potential importance for the interpretation of the supernova observations, which are still probing much smaller angular scales than the scale at which the homogeneity is recovered.
Moreover, mechanisms able to obscure lines of sight are generally related to mass concentrations, with the consequence that selection effects give a neat bias at demagnified values: in other words, they systematically hide matter from observations.
Selection effects connect indeed the meatball model to the empty or partially-filled beam formalism started by Zel'Dovich \cite{zel64} and developed later by Kantowski, Dyer, Roeder and other authors \cite{bertotti66, dashe66, gunn67, Kantowski:1969, DyerRoeder72, Weinberg:1976jq, Ellis:1998ha,Mattsson:2007tj,Bolejko:2011ej}.
We will discuss here an illustrative example where the survival probability is modeled with a simple step-function in the impact parameter.

\begin{figure}
\begin{center}
\includegraphics[height= 6.5 cm]{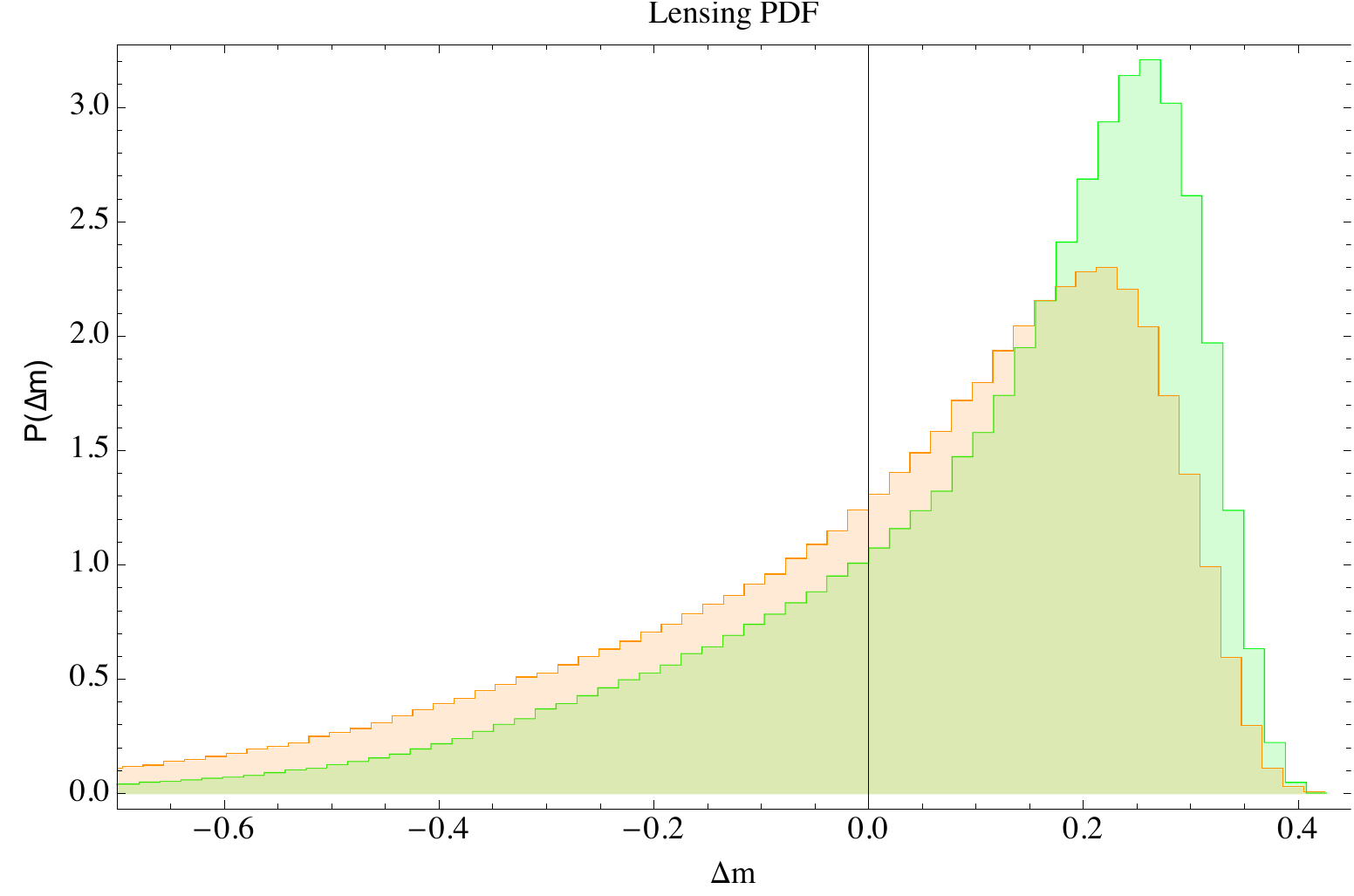}
\caption{Lensing PDFs in magnitudes for the meatball model described in Section \ref{mmo} for a source at $z=1.5$. The taller (green) histogram is without selection effects, while the shorter (orange) histogram is with selection effects. While the former has a mean at zero (background) value, the latter has a mean at the value of $\Delta m \simeq 0.09$ mag. The $y$-axis follows from the normalization $\int {\rm d}\Delta m \, {\rm P}(\Delta m)=1$.}
\label{PDFs}
\end{center}
\end{figure}

It could be possible that lensing effects due to a skewed PDF, possibly strengthened by selection effects, could explain away part of the necessary dark energy, also because lensing effects are stronger on a EdS background as compared to the concordance model.
To illustrate this point we compute the lensing PDF relative to the EdS model with $h=0.5$ (as always $H_{0}=100 \, h$ km s$^{-1}$ Mpc$^{-1}$) using the \texttt{turboGL} code \cite{turboGL} based on the sGL method~\cite{Kainulainen:2009dw, Kainulainen:2010at,Kainulainen:2011zx}.\footnote{The sGL method for computing the lensing convergence is based on generating stochastic configurations of halos and filaments along the line of sight. The \texttt{turboGL} code is a numerical implementations of the sGL method and allows to compute the full lensing PDF in a few seconds. See, for example, \cite{Holz:1997ic,Hilbert:2007ny,Hilbert:2007jd,Holz:2004xx} for lensing PDFs obtained through ray-tracing techniques.}
In this example the meatball model has two families of spherical halos, each one comprising half of the total matter density.
The halos of the first family have a mass of $10^{14} \, h^{-1} M_{\odot}$ and SIS (singular isothermal sphere) profile with radius of $R=1 \, h^{-1}$ Mpc, while the halos of the second family a mass of $10^{17} \, h^{-1} M_{\odot}$ and Gaussian profile with radius of $R=10 \, h^{-1}$ Mpc.
Moreover, we toy model the selection effects by obscuring light that hits the halos with impact parameter smaller than $20$\% of the radius.
We show in Fig.~\ref{PDFs} the lensing PDF without selection effects (taller histogram) and the PDF with selection effects (shorter histogram).
While the former has correctly a mean at zero (background) value, the latter has a mean at the (demagnified) value of $\Delta m \simeq 0.09$ mag.
Moreover, the mode is in both cases at large demagnified values, $\Delta m > 0.2$ mag.
This toy example shows, therefore, that the skewness of the lensing PDF together with selection effects may mimic the demagnifying effects of the cosmological constant.
It is also interesting to note that, once selection effects have been properly modeled, experimental data (e.g. SNe datasets) can constrain the lensing PDF and so the spectrum of matter inhomogeneities~\cite{Kainulainen:2010at}.

Lensing effects alone, however, cannot explain away dark energy and the basic reason is that they are non-negligible only for $z \gtrsim 0.5$.
Nonetheless their effect could add up to other effects.
They could, for example, enhance non-Copernican models (see next Section) or make the necessary void radius smaller \cite{Kainulainen:2009sx}.
Moreover, in the above toy model redshifts are related to comoving distances through the background metric, thus neglecting any redshift effects, which may be significant.
They may lead indeed to large corrections to the observables, especially in presence of selection effects which may change the number of blueshifted sources as compared to the redshifted ones.

\section{Non-Copernican Models} \label{noncoper}

We now examine non-Copernican models where the observer is close to the center of a very large underdensity, which have been shown to mimic accelerated expansion without dark energy.\footnote{In \cite{Celerier:2009sv, Hellaby:2009vz} it has been proposed that a model with a central overdensity can reproduce the lightcone expansion rate and matter density of the concordance $\Lambda$CDM model. However this is possible only by invoking a strongly non-simultaneous big bang, and it is not clear if this can be made compatible with other observations, such as the CMB, as we briefly discuss later in Section~\ref{SNeo}.}
We will restrict our discussion to void models based on the LTB metric and we will consider both compensated and uncompensated profiles. The former are very similar to the Swiss-cheese holes of the previous Section, the only difference being the much larger size and the central position of the observer.
Uncompensated voids, instead, are solutions that approach a background FLRW model only asymptotically.
Our discussion will be qualitatively valid for void models not based on the LTB metric, such as the so-called ``Hubble-Bubble" models, in which the inner underdensity and the outer homogenous region are simply described with two disconnected FLRW metrics~\cite{Zehavi:1998gz,Conley:2007ng,Caldwell:2007yu}.

\subsection{SNe Observations} \label{SNeo}

In the past 15 years it has been extensively studied (see, for example, \cite{Celerier:1999hp,Tomita:1999qn, Moffat:1994qy,Mustapha:1998jb,Iguchi:2001sq, Alnes:2005rw, Chung:2006xh, Enqvist:2006cg, Tanimoto:2007dq, Alexander:2007xx, GarciaBellido:2008nz, Yoo:2008su, Zibin:2008vk, February:2009pv, Sollerman:2009yu, Kolb:2009hn, Celerier:2009sv, Dunsby:2010ts, Yoo:2010qy, Biswas:2010xm, Clarkson:2010ej, Moss:2010jx,Marra:2010pg}) how an observer inside a spherical underdensity expanding faster than the background sees apparent acceleration.
This effect is easy to understand: standard candles are confined to the light cone and hence temporal changes can be replaced by spatial changes along photon geodesics. In this case, ``faster expansion now than before" is simply replaced by ``faster expansion here than there".  This is why a void model can mimic the effect of a cosmological constant if it extends to the point in space/time where the dark energy becomes subdominant: a typical scenario that can mimic the late-time acceleration of the concordance $\Lambda$CDM model consists of a deep void extending for 1-3~Gpc.

\begin{figure}
\begin{center}
\includegraphics[width= .38\textwidth, height=4.1 cm]{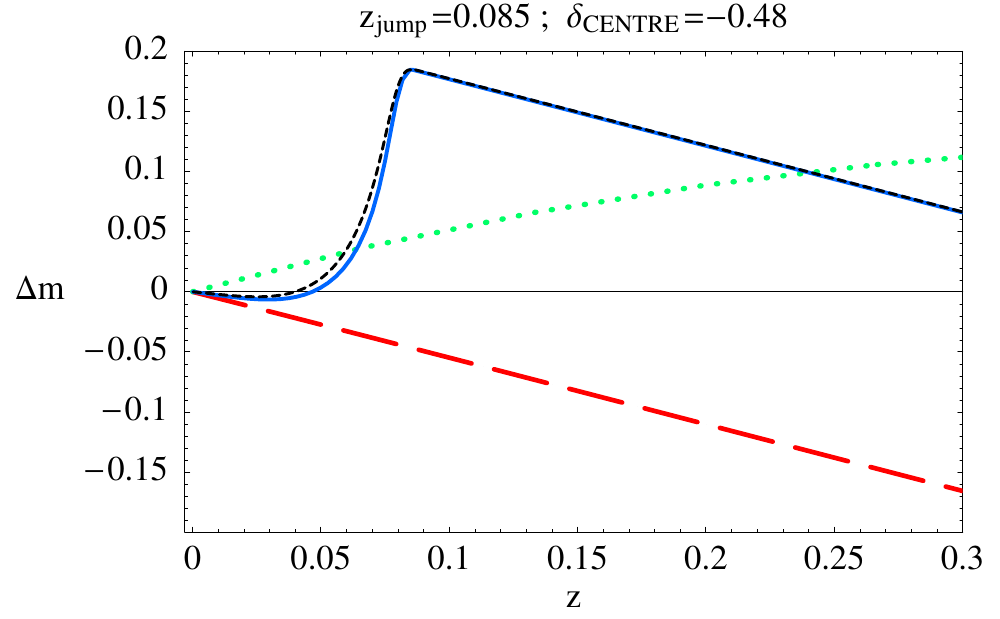}
\includegraphics[width= .54\textwidth, height=4.3 cm]{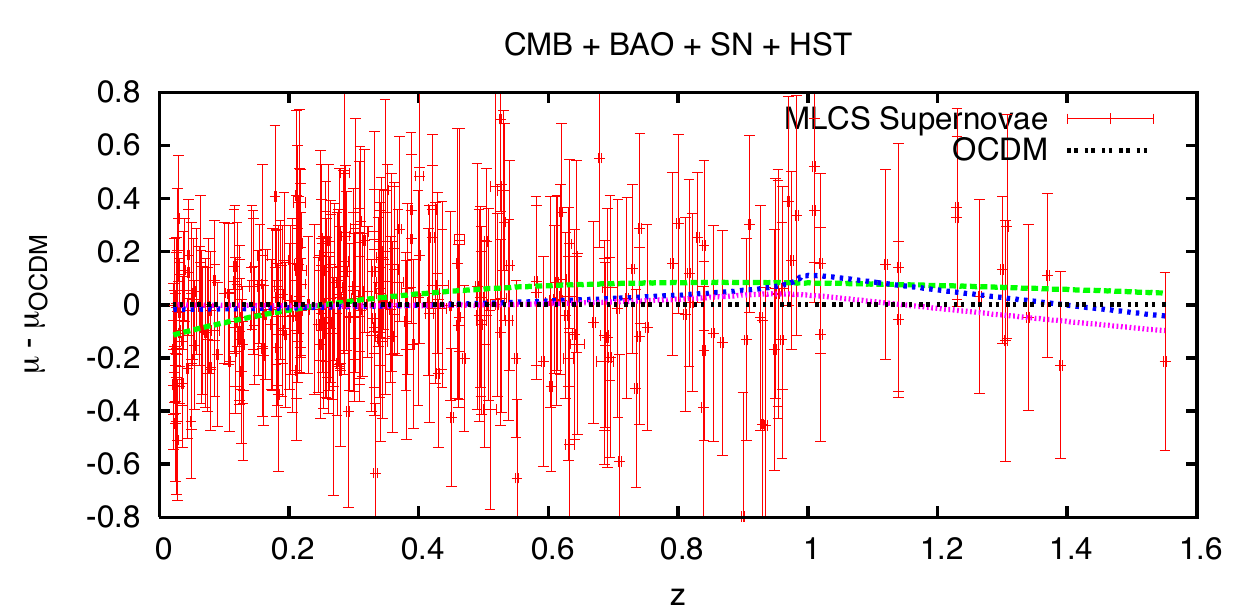}
\caption{
Left panel: comparison between analytic (short-dashed line) and numerical (solid line) $d_L(z)$ curves as obtained in~\cite{Alexander:2007xx}.
Also plotted are the EdS curve (long-dashed line) and the $\Lambda$CDM curve with $\Omega_{\Lambda}=0.7$ (dotted line).
Right panel: SN data (SDSS SN) and predictions normalized to a reference model with $\Omega_M = 0.3$ and $\Omega_K = 0.7$ (OCDM) for the best-fit model of~\cite{Biswas:2010xm} (CMB + BAO  + SNe + HST). 
Also plotted are the curves relative to  $\Lambda$CDM model (dashed green), void model with curved background (dotted blue) and void model with flat background (fine dotted magenta).}
\label{fig:CBHS}
\end{center}
\end{figure}

We may understand what happens by looking again at Eq.~(\ref{dez}). In a void model slightly smaller than the horizon the dominant term in the Newtonian gauge is the Doppler effect, which induces a large net correction to the redshift and modifies the $d_L(z)$ curve in the same way of an accelerating universe; in other words, it is as if all sources in the underdense region had a radial collective peculiar velocity which gives them an extra redshift.
This is the dominant effect at low redshift and therefore the most relevant for SNe. This behaviour has been extensively studied through analytical approximations and both in the synchronous and Newtonian gauge in~\cite{Biswas:2006ub,Biswas:2007gi,Alexander:2007xx}, an example of which is given in the left panel of Fig.~\ref{fig:CBHS}. In addition there is a correction to the distance itself, which, however, becomes important only at high $z$:  a change in the angular metric element $Y(r,t)$ induces lensing corrections to the area under which a single source is seen.

More formally one may consider the freedom available within LTB models: as explained in \ref{LTB}, the LTB metric features indeed two free functions. In the usual description the LTB model is analyzed in comoving and synchronous coordinates, where one free function called the ``bang time" function $t_B(r)$ sets the time of the big bang at each value of the coordinate $r$, and the other function can be taken as the density profile or the $r$-dependent curvature (depending on arbitrary redefinitions of the radial coordinate, which leave the LTB metric in the same form).
It is enough to adjust one free function (e.g., the curvature) to obtain any distance-redshift relation,\footnote{Strictly speaking it is not possible to have a negative deceleration parameter $q_{0}$ at the coordinate point of the observer, unless one uses a density profile which is spiked at the origin~\cite{Vanderveld:2006rb,Krasinski:2009qq}. Nonetheless, it is easy to evade this constraint observationally with a rapidly varying function or by smoothing the density profile on a appropriately small scale~\cite{Yoo:2008su}, since it is true only in one point. This fact, however, might be of some observational interest for future precise measurements at very low redshifts.} as stated by Ref.~\cite{Mustapha:1998jb} and illustrated, for example, by Refs.~\cite{Yoo:2008su,Kolb:2009hn} where the luminosity distance of the concordance model is reproduced (see also \cite{Romano:2009mr}).
By adjusting the other free function, it is possible to obtain also the light-cone matter density (or galaxy number count) of the concordance model \cite{Kolb:2009hn}.
However, it is probably preferable  to avoid an inhomogeneous bang function (see the simultaneous big bang condition in Eq.~(\ref{age})),
which would introduce very large inhomogeneities in the past, strongly at odds with the inflationary paradigm~\cite{Biswas:2006ub,Zibin:2008vj}. It is possible, however, to demand that $t'_B(r)\approx 0$, but not strictly zero.
In this case one should guarantee that observations are compatible with the inhomogeneities in the big bang time. In particular, constraints can be imposed by demanding that, for a radius $r_{LSS}$ and time $t_{LSS}$ which correspond to the Last Scattering Surface, the  density perturbations associated to $t'_B(r)$ are small, in order not to spoil CMB observations~\cite{Bolejko:2004vb}. This implies a strong suppression at present time on $t'_{B}$ at $r\approx r_{LSS}$, since these are decaying modes. More freedom could be allowed for $r \ll r_{LSS}$, keeping in mind however that one should not introduce too large secondary effects on the CMB, and not spoil galaxy number counts and galaxy ages.
In the rest of the review we will always set the bang function to $t_B=0$, which is the simplest choice and also makes the model more constrained with respect to the $t'_B\approx 0$ case.

We show in the right panel of Fig.~\ref{fig:CBHS} an example from~\cite{Biswas:2010xm} of how SNe data can be fitted with a void model.
The top left panel of Fig.~\ref{figLLTB} illustrates the fact that concordance and void model equally well fit the SNe distance-redshift diagram, showing confidence level contours on the void radius $r_{0}$ and $\Omega_{\Lambda, \textrm{\scriptsize out}}$ for the $\Lambda$LTB model of Ref.~\cite{Marra:2010pg} (the subscript ``out'' labels quantities relative to the background model).
In \cite{Marra:2010pg} the void has the smooth density profile already used in \cite{GarciaBellido:2008nz}, which depends crucially only on the void radius and depth, the two main physical quantities describing an underdensity.

\subsection{CMB} \label{seCMB}

\begin{figure}
\begin{center}
\includegraphics[width= .48\textwidth]{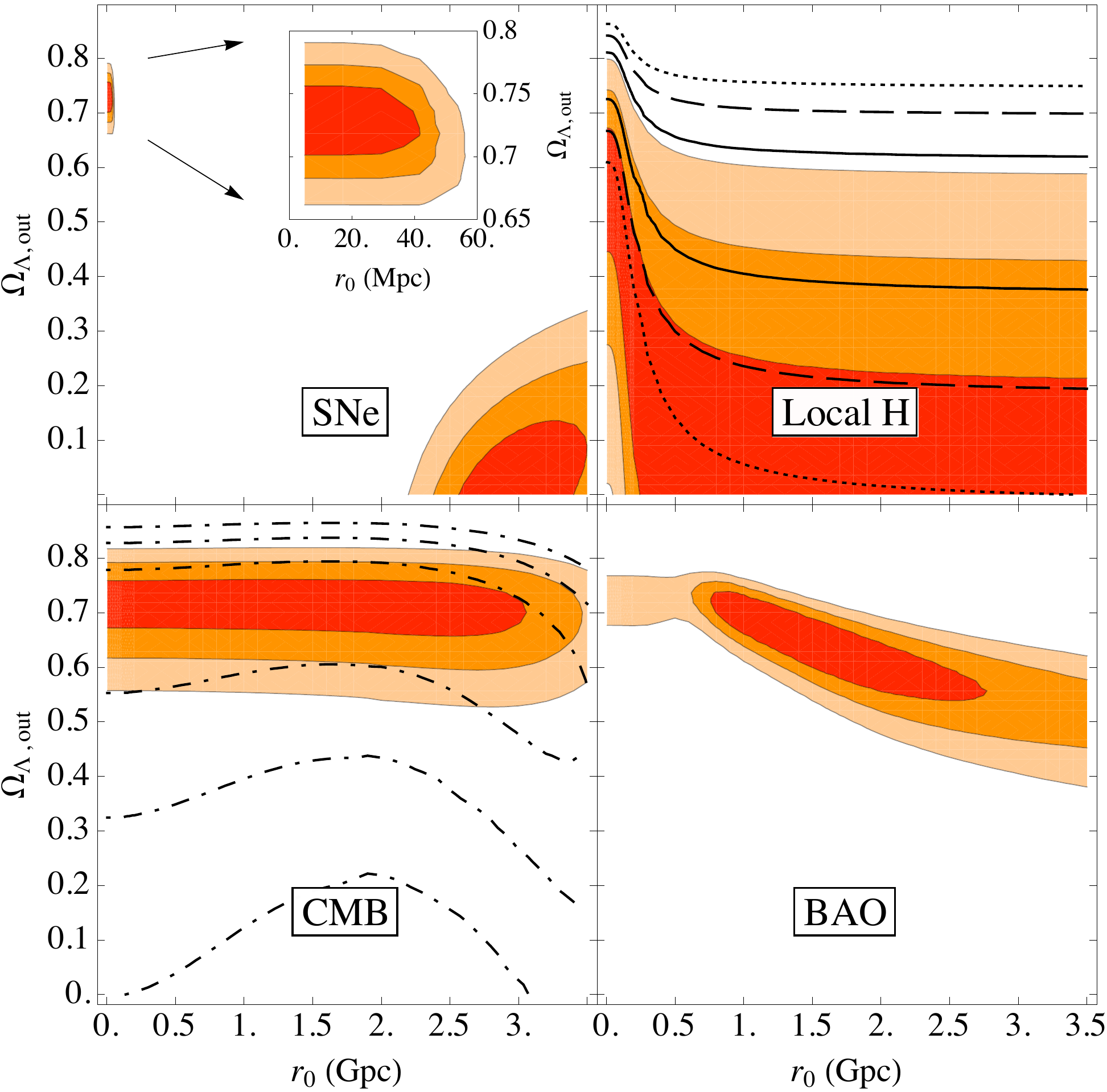}
\includegraphics[width= .48\textwidth]{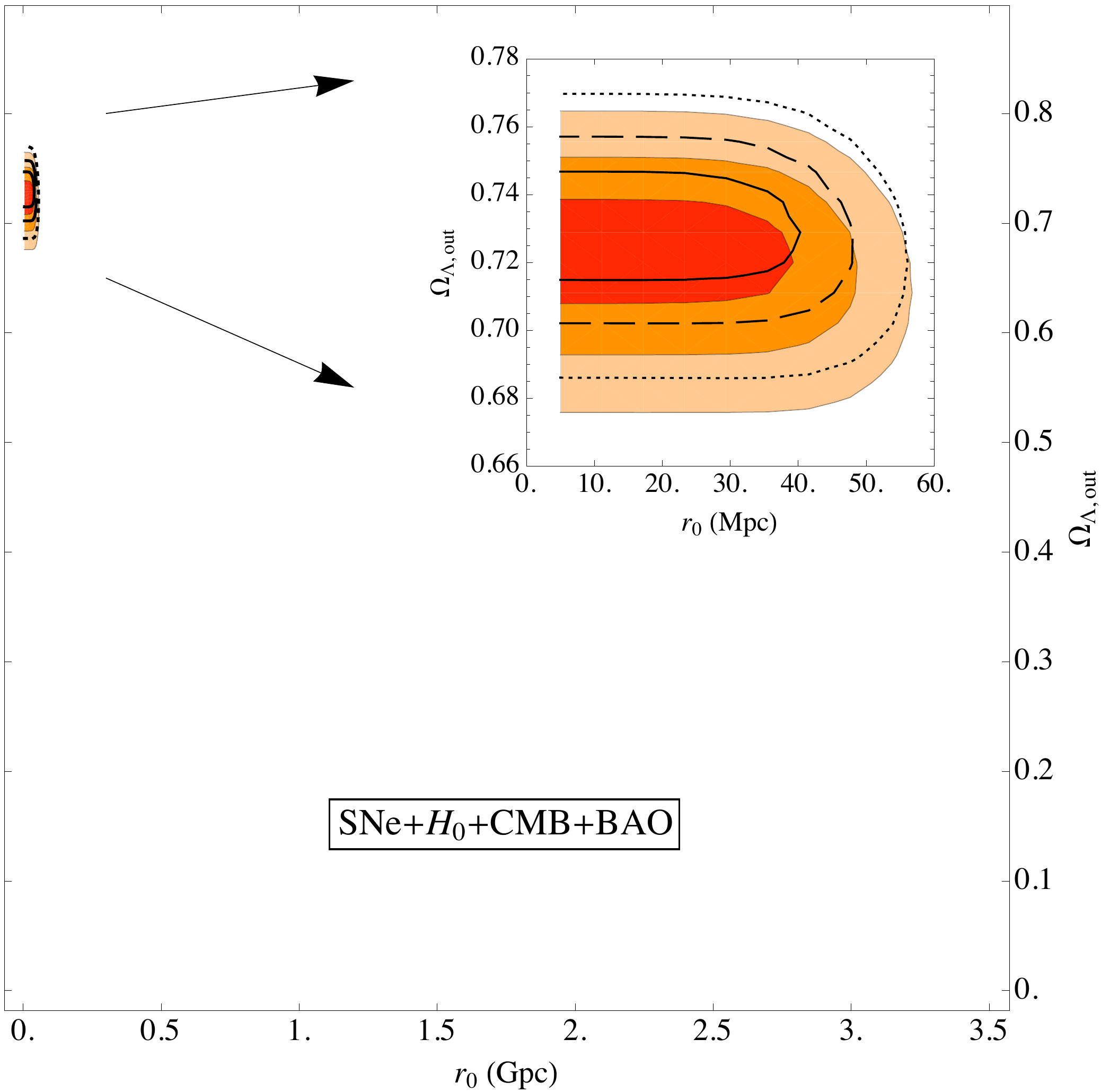}
\caption{
1, 2 and 3$\sigma$ confidence level contours for the asymptotically flat $\Lambda$LTB model of~\cite{Marra:2010pg} on the void radius $r_{0}$ and $\Omega_{\Lambda, \textrm{\scriptsize out}}$  with $n_{s}=0.96$ and $t_{0}=13.7$ Gyr.
The four smaller panels on the left show the contours for the independent likelihoods per observable, while the larger panel on the right shows the contours for the combined likelihood.
In the panel representing measurements of the local Hubble constant, the results relative to $H_{\textrm{\scriptsize S06}}$ of Eq.~(\ref{S06}) are shown as filled contours, while the ones relative to $H_{\textrm{\scriptsize R09}}$ of Eq.~(\ref{R09}) are shown as lines. The same labelling holds for the panel relative to the combined observables.
In the panel relative to CMB constraints we also show confidence levels for the likelihood marginalized over $n_{s}$ (dot-dashed contours).
In the panels, the $x$-axis represents pure-matter void models, while the $y$-axis the standard FLRW model.
}
\label{figLLTB}
\end{center}
\end{figure}

Dust LTB void models are matched to the background metric at a redshift at which radiation is still negligible; a value of $z \sim 100$ satisfies, for example, this requirement.
In this way the Last Scattering Surface (LSS), which is responsible for most of the CMB anisotropies, is outside the inhomogeneous patch (and so there are no Doppler effects) and a standard analysis of the primordial CMB power spectrum is possible (but perhaps not of the BAO as we will discuss in the next Section).
Moreover, the lensing effects discussed in the previous Section are almost vanishing, as light rays always go radially for a central observer in spherically symmetric void models\footnote{Light propagation is qualitatively different for the non-spherically symmetric setup of Fig.~\ref{schizzo}.} and the angular diameter distance $d_{A}(z) = Y(r(z),t(z))$ is very close to the one of the background model: the scale factor $Y$ exactly matches the background one ($Y= a \, r$ for $r> r_{0}$) and the only difference is in a small correction (redshift effect) to the arguments $r(z)$ and $t(z)$ when solving for the geodesics. 
This small shift in the geodesics slightly changes the distance to the LSS and it can be reabsorbed into a redefinition of the photon temperature of the CMB.
In other words, everything can be analyzed with an effective FLRW model with a slightly different value of $T_0$~\cite{Zibin:2008vk,Biswas:2010xm}, which will generally differ from the usual $2.726$ K and consequently, the LSS is located at a slightly different redshift. Note also that there are other contributions to the CMB coming from secondary effects, which may differ from FLRW and are due to the photons traveling through the inhomogeneities inside the void, such as the ISW effect or lensing; in the literature these effects have not been considered, since they are subdominant and since studying them would require knowledge of the growth of perturbations in an LTB model, which is not yet completely understood (see, however, \cite{Zibin:2008vj,Clarkson:2009sc}). 

Now, in a compensated model the correction to $T_0$ is of order $(l_{0}/l_{\rm hor})^2$~\cite{Biswas:2010xm}, therefore small for void radiuses of about 1-3 Gpc (${\cal O}(1\%)$, ~\cite{Zibin:2008vk, Biswas:2010xm,Marra:2010pg} ), and since it is well-known that it is not possible to fit the CMB observations with an EdS model, it follows that (assuming the standard primordial power spectrum) compensated void models which are
asymptotically flat are ruled out.
This holds independently of how the LTB free functions are adjusted and is shown by the bottom left panel of Fig.~\ref{figLLTB} taken from~\cite{Marra:2010pg}. Similarly, the asymptotically flat void model has been shown by~\cite{Biswas:2010xm} to be excluded with a $\Delta\chi^2\approx 35$.
In~\cite{Marra:2010pg} the CMB power spectrum has been analysed using accurate fits for the positions of the first three peaks and the first trough and the relative heights of the second and third peak relative to the first one, while in \cite{Biswas:2010xm} the full $C_{l}$ spectrum has been studied using a modified version of \texttt{COSMOMC}. The agreement between the two approaches strengthens their conclusions.

One way of improving the fits is to consider uncompensated voids, which have a large correction to the monopole temperature~$T_0$ and therefore a different distance to the LSS.
It has been shown~\cite{Zibin:2008vk, Biswas:2010xm} in fact that such a model with an ${\cal O}(20\%-40\%)$ correction to the monopole, taken at face value, can have a much better fit than the asymptotically flat model.
It is not clear, however, if in this case it is consistent to analyze the CMB in the standard way, since the LSS itself is affected by a radial peculiar velocity, and so these models may have to be considered only illustrative of the potential effects they cause to the observables~\cite{Biswas:2010xm}.

A simpler way out of the shortcomings of the asymptotically flat model is to consider void models that are asymptotically curved \cite{Biswas:2010xm, Marra:2010pg} and so also with a different distance to the LSS: a closed model (e.g.,~$\Omega_K \approx  -0.2$) older than the $\Lambda$CDM concordance model (e.g.,~$t_0 \approx 16 \div 18$ Gyr) can indeed fit the observed CMB, as shown by Table~\ref{chi2tabb} which corresponds to the profiles of Fig.~\ref{BNV} studied in Ref.~\cite{Biswas:2010xm}.
Note that these fits include also the value of $H_0$ and BAO constraints, as discussed in the next two Sections.
We can summarize the previous discussion by stating that CMB observations fix the background metric of a compensated void model, with a weak dependence on the LTB free functions.

\subsection{Local $H_{0}$}

It is well known that a curved FLRW model can fit the CMB only at the price of a very low value of the Hubble rate and this remains true, as discussed in the previous Section, for the Hubble parameter $H_{\rm out}$ of the FLRW region outside the void.
When comparing with observations, however, the presence of the void itself alleviates the problem as the observer experiences a higher local expansion rate $H_0$.
The jump $\Delta H=100 \, \Delta h$ km s$^{-1}$ Mpc$^{-1}$ between $H_{\rm out}$ and $H_0$ is constrained by a good fit to the SNe to a value of  $\Delta h \sim 0.2$ and so we can conclude that CMB observations fix the magnitude of the local Hubble rate, which can then be confronted with the observations. We will consider the following two results from Ref.~\cite{Sandage:2006cv} and Ref.~\cite{Riess:2009pu}:
\begin{eqnarray}
H_{\textrm{\scriptsize S06}}&=&62.3 \pm 5.2  \textrm{ km s}^{-1} \textrm{Mpc}^{-1} \phantom{ciaooo} \textrm{Sandage et al. 2006}  \,,  \label{S06} \\
H_{\textrm{\scriptsize R09}}&=&74.2 \pm 3.6  \textrm{ km s}^{-1} \textrm{Mpc}^{-1}  \phantom{ciaooo} \textrm{Riess et al. 2009} \,.  \label{R09} 
\end{eqnarray}
The general trend is that the CMB favors values of the local expansion rate which are at most of about $H_0\approx 50   \textrm{ km s}^{-1} \textrm{Mpc}^{-1} $~\cite{Biswas:2010xm, Marra:2010pg}, so in disagreement with $H_{\textrm{\scriptsize R09}}$, but in marginal agreement with $H_{\textrm{\scriptsize S06}}$.
It is, however, possible to exploit further the freedom of the LTB models in order to increase the local value of $H_0$ by adjusting the LTB free functions in order to have a small local patch around the observer, with a higher expansion rate, that does not disrupt the SNe fit \cite{Biswas:2010xm} and this has been shown to also give a better overall combined fit to cosmological observables, as can be seen in Table~\ref{chi2tabb}.

Another possibility considered in the literature, alternative to the simplest asymptotically flat models, is to adopt a non-standard (but physically motivated) primordial power spectrum as suggested by Ref.~\cite{Nadathur:2010zm} (see also \cite{Moss:2010jx}), where a void model matched to EdS was shown to be in agreement with SNe, CMB and $H_{0}$ observations.
Yet another possibility is to consider an inhomogeneous profile for the radiation~\cite{Clarkson:2010ej}, thus allowing for the extra freedom necessary to accommodate the CMB power spectrum.

\begin{table}
\begin{tabular*}{\textwidth}{@{\extracolsep{\fill}}  l|rrrr|r}
Model & CMB & BAO & SNe & $H_{\textrm{\scriptsize S06}}$ & Total $\chi^2$\\
\hline
$\Lambda$CDM				& 3372.1 & 3.2 & 239.3 & 0.4 &{\bf 3615.0}\\
Profile A (Curved Void)  	& 3377.1 & 4.3 & 240.7 & 6.6 &     3628.7\\
Profile B						& 3377.2 & 0.6 & 235.3 & 5.1 &     3618.2\\
Profile C						& 3376.9 & 1.0 & 234.9 & 3.7 &{\bf 3616.5}\\
Profile D						& 3376.7 & 3.8 & 233.9 & 2.2 &     3616.6\\
Profile E						& 3372.9 & 3.2 & 241.9 & 1.1 &     3619.1
\end{tabular*}
\caption{A breakdown of the total $\chi^2$ for each dataset, for fitting simultaneously to CMB + BAO + SNe + $H_{\textrm{\scriptsize S06}}$.
The profiles refer to Fig.~\ref{BNV} and were studied in Ref.~\cite{Biswas:2010xm}.\label{chi2tabb}}
\end{table}

\subsection{BAO and Power Spectrum}

\begin{figure}
\begin{center}
\includegraphics[height= 6 cm]{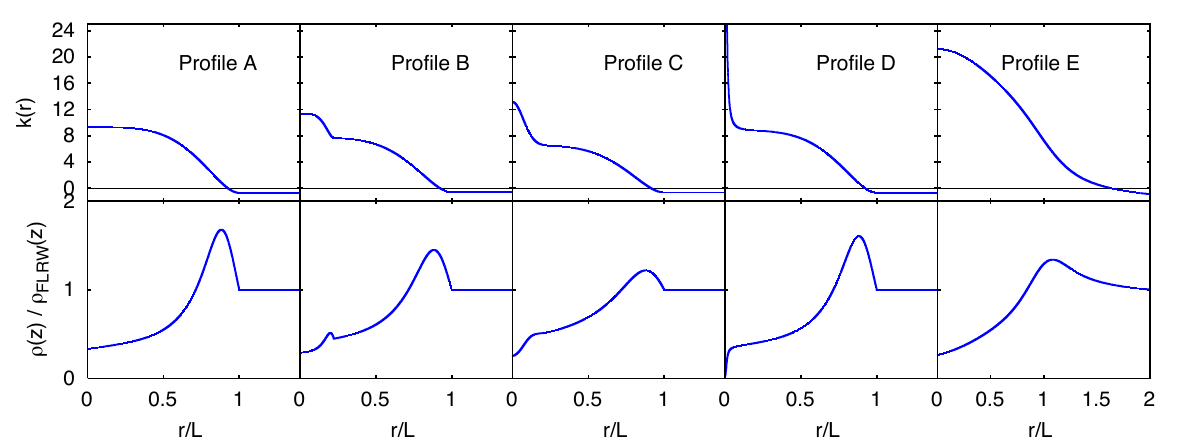}
\caption{Curvature (top) and density (bottom) void profiles studied by Ref.~\cite{Biswas:2010xm}, whose corresponding $\chi^2$ are displayed in Table \ref{chi2tabb}.
By modeling the void with the profiles B and C it is possible to obtain an excellent fit to the BAO observations.}
\label{BNV}
\end{center}
\end{figure}

A baryon acoustic peak has been detected from SDSS and 2dFGRS galaxy catalogues at a mean redshift of $0.2$ and $0.35$ \cite{Percival:2007yw}.
This measurement constrains theoretical models by means of the following ratio:
\begin{equation}
\theta(z)\equiv {L_S\over d_V(z)} \, ,
\label{theta}
\end{equation}
where $L_S$ is the comoving sound horizon scale at recombination and $d_V$ is a combination of angular and radial distance defined as follows:
\begin{equation}
d_V=[(1+z)^2 \, d_A^2 \, d_z]^{1/3} \, .
\end{equation}

Void models able to fit the SNe have to extend at least up to $z\sim 1$ and so the BAO feature is well inside the inhomogeneous patch, where spatial gradients are strong.
It is possible to compute in a given LTB model~\cite{Biswas:2010xm} the above distances, taking into account that the transverse and the radial expansion rates are different.
However, in order to  compare with observations one needs to make assumptions, which make the  validity of the analysis unclear.
The first assumption is to ignore the effects of the background on the growth of perturbations, which is unlikely to be correct in an LTB model because of the anisotropies in the expansion rate.
As for the secondary effects on the CMB discussed in Section \ref{seCMB}, this has not yet been addressed in the literature since the growth of perturbations in an LTB model is not clearly understood (see Ref.~\cite{Zibin:2008vj,Clarkson:2009sc}).
The second assumption is to consider that $L_S$ is given by FLRW recombination physics, which is unlikely to be correct in this kind of models: it is true that  the baryon acoustic scale is imprinted in the sky at early times, when the universe was supposed to be more homogenous, but it has been imprinted in a spatial location in which the void itself was going to develop and therefore this should be analyzed in a radiation dominated era within a spherically symmetric background.
Further, in order to perform a more accurate analysis one should compute $\theta(z)$ for a generic $z$ (and not only for the two values above) and then weight it, for instance, by the corresponding number density of observed objects, especially if the void profile is rapidly varying in this redshift range~\cite{Biswas:2010xm}.

If one insists, with the above caveats,  in confronting LTB models against BAO observations, one finds that a flat (near the origin) density or curvature profile generally has difficulties in fitting the BAO scale, as shown in the example of Fig.~\ref{figLLTB}.
It is however possible by a more specified tuning of the void profile to fit the BAO data without cosmological constant. A better fit than the $\Lambda$CDM model can be found indeed by adopting the profiles shown in Fig.~\ref{BNV}, as can be seen in Table~\ref{chi2tabb} from~\cite{Biswas:2010xm}.

Similar considerations apply to other constraints coming from the matter power spectrum.
Here also it is essential to study the growth of perturbations in an LTB model, and again this has not yet been done (see however~\cite{Alonso:2010zv} for recent improvements using numerical simulations). A fit of these data can nonetheless be performed under some rough assumptions, in order to get some qualitative indication.
In~\cite{Biswas:2010xm}, for example, fits have been performed assuming the growth of structures as in an approximate effective FLRW model built up from the parameters at the centre of the void. The results show that in this case the void gives a bad fit, be it with $\Omega_{K, {\rm out}}=0$ or with $\Omega_{K, {\rm out}}\neq0$, since matter power spectra favour a value for the combination $\Omega_{\rm DM, in} h$ of about $0.2$, while the curved void  from the fit with CMB + BAO + SNe + $H_{\textrm{\scriptsize S06}}$ has a best fit value of about $0.09$. Note, however, that the fit is significantly better for the LRG subset~\cite{Tegmark:2006az} than for the SDSS main sample~\cite{Tegmark:2003uf}.  In fact,  the void does predict too much power on large scales and the LRG data prefer this, as compared to the SDSS data.

\subsection{Constraints on the observer's position.}
\label{secobserver}

Most works in the literature take the observer to be at the center of the LTB void in order to recover the observed isotropy of the universe.
This is undoubtedly a heavily fine-tuned configuration, especially for the very large voids considered.
It is therefore interesting to see which actual limits are placed on the off-center position of the observer by present or future observational data.

Using the anisotropies in the luminosity of the observed supernovae (see~\cite{Schwarz:2007wf} for a claim of detection of discrepancy between the equatorial North and South hemispheres), it has been found \cite{Alnes:2006uk, Blomqvist:2009ps} that SNe observations constrain the observer's position to about $200$ Mpc from the void center.
Much tighter constraints come, however, from the observed dipole of the CMB; this is best understood in the Newtonian picture, see \cite{Alnes:2006pf} for exact numerical results and \cite{Kodama:2010gr} for analytical results valid for general spherically-symmetric spacetimes.
An observer close to the center and comoving with the LTB metric has a ``background''  peculiar velocity with respect to the background model (and the void center) of $\Delta v = \Delta H \, d_{\rm obs}$, which is completely due to the inhomogeneous expansion rate (as opposed to the ``real'' peculiar velocities in the standard FLRW paradigm; see Ref.~\cite{Kolb:2009rp} for a discussion about peculiar velocities in inhomogeneous universes).
This causes a Doppler shift in the CMB temperature and a consequent dipole of magnitude~\cite{Alnes:2006pf}:
\begin{equation} \label{dipole}
a_{10}  = \sqrt{\frac{4\pi}{3}}   \, {\Delta v \over c}  = \sqrt{\frac{4\pi}{3}}  \, \frac{\Delta h}{3000 \, \textrm{Mpc}} \, d_{\rm obs} \,.
\end{equation}
As said earlier, SNe data typically constrain void models to have $\Delta h \sim 0.2$ and so the dipole of Eq.~(\ref{dipole}) has a magnitude of the order of the observed one for an observer displaced from the center of about 20-30 Mpc.
For smaller displacements, the LTB-induced dipole is smaller and peculiar velocities are invoked, as in the FLRW model, to match the observed value.
For larger displacements, the induced dipole is too big and the observer's background peculiar velocity has to be compensated by a real peculiar velocity directed towards the center of the void.
It has been found by \cite{Quartin:2009xr,Foreman:2010uj} that, for typical values of real peculiar velocities, the position of the observer is constrained to be up to 60--80 Mpc from the center, which is about few percents of the typical radius of a void model.
The fine tuning on the observer's position is therefore heavy.
It is worth mentioning, however, that such a tuned position of the observer could simultaneously explain \cite{Biswas:2010xm} the recent observation of large bulk flow velocities~\cite{Kashlinsky:2008ut} of order $600\div 1000$ km/sec in a local region of about $300 h^{-1}$ Mpc close to the direction of the observed CMB dipole.
An observer displaced from the center would indeed observe a difference in velocities in the sky with a dipolar shape aligned with our local CMB dipole, with a magnitude of the order of $\Delta v$ in agreement with \cite{Kashlinsky:2008ut}.

Finally, it has been shown by \cite{Quartin:2009xr,Quercellini:2008ty} that tighter constraints will come from future high-precision astrometric observations of distant quasars.
Cosmic parallax measurements can indeed put stringent bounds ($\sim$10 Mpc) on the off-center position of the observer, which are independent from the constraints relative to the CMB dipole.
In particular, the effect depends on the source and observer peculiar velocities,  and this may break the degeneracy between real and background peculiar velocities discussed above.
Cosmic parallax together with the redshift drift (to be discussed in the next Section) belong to the realm of the so-called ``real-time'' cosmology~\cite{Quercellini:2010zr}.

It is important to stress that the previous discussion is relative to the LTB metric in which only one center position can be meaningfully defined.
All the previous constraints could indeed be in principle alleviated if one considers non spherically-symmetric metrics which allow for more (or none) center positions~\cite{Bolejko:2010wc}.
In other words, in more general metrics the center location of the LTB models could become an extended volume, thus reducing the fine tuning required on the observer's position.  
It would be  interesting to quantitatively define the amount of fine tuning within a void model by taking, for example, the ratio of the volume occupied by observers who see a CMB dipole smaller than or equal to the observed one to the total volume of the void region.
In this way one could precisely see how weakened is the fine tuning on the observer's position in more general geometries as compared to the LTB case.

\subsection{General tests of the Copernican Principle}

We will now discuss more general tests and observables able to further constrain the void models, which apply also when the observer is exactly at the center.


Even if we do live at the center of the LTB void, other halos and clusters in the universe do not and the corresponding observers will see a large dipole in the CMB.
Such a dipole would manifest to us observationally through the kinematic Sunyaev-Zel'Dovich (kSZ) effect.
The hot electrons inside a cluster distort indeed the CMB spectrum through inverse Compton scattering, in which the low energy CMB photons receive energy boosts during collisions with the high-energy cluster electrons.
In the (first-order) thermal Sunyaev-Zel'Dovich effect the CMB photons interact with electrons that have high energies due to their temperature, while in the (second-order) kinematic Sunyaev-Zel'Dovich effect the CMB photons interact with electrons that have high energies due to their bulk velocity.
The kSZ effect is proportional to the radial velocity of the cluster with respect to the CMB scattering surface and can be employed to map the cosmic peculiar velocity field inside our own light cone.

Ref.~\cite{GarciaBellido:2008gd} by examining available kSZ measurements was able to put bounds on the size of the void radius, which is then constrained to be smaller than about 1.5 Gpc.
They also conclude that, within their adopted void modeling, the kSZ constraints cannot be satisfied if the simultaneous big-bang conditions is demanded.
This clearly shows how tests beyond the observer's lightcone are crucial to constrain void models (see also \cite{Garfinkle:2009uf}).
Moreover, it would be interesting to quantify the amount of inhomogeneity in the bang function (and so of decaying modes) that is necessary to fit the kSZ data.
Ref.~\cite{Yoo:2010ad} examines the same dataset considered by Ref.~\cite{GarciaBellido:2008gd} and finds their void model ruled out due to the high peculiar velocities. The different conclusion with respect to Ref.~\cite{GarciaBellido:2008gd} might be attributed to the different modeling of the void; see indeed Fig.~3 of~\cite{GarciaBellido:2008gd} and Fig.~4 of~\cite{Yoo:2010ad}, which show that the two models have very different radial velocities at $z\gtrsim 0.5$.
Ref.~\cite{Yoo:2010ad} considers also the case of an inhomogeneous decoupling hypersurface.
A large kSZ effect is indeed due to a large velocity between the comoving clusters and the CMB frame and this can be alleviated if radial non-adiabatic (isocurvature) inhomogeneities in the non-relativistic matter on the decoupling hypersurface are introduced.
More work is necessary to understand the implications of such a non-standard universe (see also \cite{Clarkson:2010ej}).

Strong results have been obtained recently by Ref.~\cite{Zhang:2010fa} where the kSZ effect due to all free electrons was calculated and confronted with the limits coming from the measurements of the small scale anisotropy power spectrum by the Atacama cosmology telescope. According to Ref.~\cite{Zhang:2010fa} void models are robustly excluded.
This test is worth careful consideration, in order to conclude on the possible definitive failure of the void models, examining in detail its assumptions and calculations (see Ref.~\cite{zibinFOCUS} of this Special Issue). For instance, one caveat is that it is assumed that the growth of perturbations is well described by a  $\Lambda$CDM model and the validity of this assumption is unclear. Moreover, non-adiabatic models could also help in avoiding these constraints.

Other tests of void models (and so of the Copernican Principle) have been considered in the literature \cite{Goodman:1995dt,Clarkson:2007pz,Clifton:2008hv}.
Rescattering of photons by off-center reionized structures can distort, for example, the CMB blackbody spectrum via Compton y-distortion~\cite{Caldwell:2007yu,Moss:2010jx}.


Another potentially interesting observable is the redshift drift, namely the temporal variation of the redshift of distant sources like quasars as a tracer of the background cosmological expansion, which was first discussed by Sandage in 1962 \cite{sandage62} (see also \cite{Loeb:1998bu,Corasaniti:2007bg}).
It relies on high-precision spectroscopy and the necessary statistical sensitivity could be reached with the next generation of optical telescopes. In particular, a redshift drift measurement $\Delta_{t}z$ over a time-span $\Delta t$ could distinguish between the real acceleration driven by dark energy ($\Delta_{t}z>0$) and the apparent acceleration of the void models ($\Delta_{t}z<0$).
Most importantly, it is a direct measurement of the local expansion rate of the universe which is independent from the evidence for acceleration given by the SNe, thus being an important test independent of the calibration of standard candles and the relative uncertainties.
The redshift drift in relation to LTB void models has been examined, for example, in \cite{Dunsby:2010ts,Quartin:2009xr,Uzan:2008qp, Yoo:2010hi}.

Finally, one can further constrain the LTB void models by means of other observational quantities such as source number counts~\cite{Romano:2009mr} and, more recently, galaxy ages~\cite{Bolejko:2011ys}.

\section{Discussion and outlook} \label{conclusions}

In this paper we have focused on the modeling of voids, which are the characterizing feature of the late inhomogeneous universe.
We have in particular discussed the possibility that the observed dark energy could be explained away by the effect of large-scale nonlinear inhomogeneities. 
We considered both the Copernican and non-Copernican paradigm and we discussed how cosmological observations can constrain these models.
Work still has to be done in order to thoroughly explore inhomogeneous models, but it is nonetheless worth drawing some partial conclusions and maybe indicate possible future directions.

Within Copernican models it is useful to distinguish between two {\em non-exclusive} effects of large-scale inhomogeneities on cosmological observables.
The first one addresses the question of how the cosmological background reacts to the nonlinear structure formation, while the second focuses on the propagation of photons in the inhomogeneous universe.
The former effect has been called sometimes ``strong'' backreaction and the latter ``weak'' backreaction (see Ref.~\cite{Kolb:2009rp} for the precise definitions).
Strong backreaction has been discussed in other contributions to this Special Issue~\cite{kolbFOCUS, buchertFOCUS, ellisFOCUS, chrisFOCUS, wiltshireFOCUS, rasanenFOCUS, phasespaceFOCUS}, but we can nonetheless say that a common agreement on its magnitude has not yet been reached.
The models reviewed in this paper cannot answer this question because by construction the evolution of the background is unaffected by the inhomogeneities introduced, despite them being fully nonlinear.
On the other hand we have studied photon propagation in these models and, from what we have presented here, it seems that a single (weak backreaction) effect on photons is not sufficient to have a paradigm shift in which dark energy is no longer needed~\cite{Marra:2007pm, Marra:2007gc,Biswas:2006ub,Biswas:2007gi}.
We think, however, that Copernican models should be studied in more detail, because even if each single effect is small, of order 1\%-10\%, their sum might not be, especially taking into account selection effects. This conjecture is strengthened by the fact that different effects of inhomogeneities seem to pull into the same direction with less and less need for dark energy~\cite{Kainulainen:2009sx}: to conclude on their viability as the possible explanation of the observed acceleration, it is therefore crucial to consider all possible effects together in as realistic a model as possible.

With non-Copernican models the situation is, to some extent, opposite.
Since the very first works of 10-15 years ago it was indeed possible to mimic the dark energy, and the issue of the viability of these models was more philosophical than observational.
However, since then, all the research has focused on exploiting the freedom available in order to accommodate more observables and recent data.
Most literature has focused on models which employ only one of the two free functions in LTB, setting the other by asking a homogenous bang function (relaxing this assumption to an almost simultaneous big bang could make it easier to satisfy the constraints).
With more data and more research it appears that for one single free function, it is by now not sufficient to consider the simplest void models embedded in EdS to fit all data, but it is necessary to introduce some features, like overall background curvature, more elaborated profiles, or nontrivial primordial spectra for the CMB. 
In this way, at least when considering CMB+SNe+$H_0$, such models succeed in fitting the data with more or less success, depending on the specific profiles~\cite{Biswas:2010xm,Marra:2010pg}.
The inclusion of the BAO observations can be done only at a qualitative level, as it would require knowledge of evolution of perturbations in LTB during the radiation and matter eras; the present analyses have done under simplifying assumptions and indicate that these data could be also fit by LTB, especially with peculiar shapes of the density profile. Similarly, one may also attempt a fit of  the matter power spectrum under the same assumptions, while the result in this case is that the fit is significantly worse~\cite{Biswas:2010xm}.

Finally, a significant and important amount of work has shown that these models can be severely constrained using the fact that there are large radial peculiar velocities. The position of the observer is indeed constrained to be at a distance of at most about 60-80 Mpc from the center of the void, in order to avoid a too large dipole in the CMB. This represents a strong fine-tuning on the observer's position (which perhaps could be weakened by modeling the void with a less-symmetric metric than LTB,  as we briefly discuss in~\ref{secobserver}), but on the other hand an off-center observer could explain the recently claimed large bulk flows on very large scales~\cite{Kashlinsky:2008ut}.
A second very interesting set of constraints comes from the fact that light from CMB  is rescattered by electrons which live in structures with a large radial collective velocity. These tests, which apply also if the observer is exactly at the centre, put severe constraints
on such models and may turn out to  rule out most, if not all, of the void scenarios.
Some of these constraints rely again on simplifying assumptions on the  growth of perturbations, and therefore more refined analyses might be needed to conclude on the viability of non-Copernican models.

\ack

We thank Lars Andersson and Alan Coley for organizing this interesting special focus issue on Inhomogeneous Cosmological Models and Averaging in Cosmology.
We also thank Luca Amendola for clarifications about the real-time cosmology and Wessel Valkenburg for clarifying discussions about Swiss-cheese and void models.
We benefited from discussions with Krzysztof Bolejko and Syksy R\"as\"anen.

\appendix

\section{LTB basic formalism} \label{LTB}

We will now quickly review the conventional LTB formalism.
For more details see, for example, Ref.~\cite{Biswas:2006ub,Biswas:2007gi,Enqvist:2006cg,Sussman:2011na}.
The line element of the spherically symmetric LTB model in comoving and synchronous coordinates can be written as ($c=1$):
\begin{equation} \label{metric}
   \textrm{d}s^2=-\textrm{d}t^2+\frac{Y'^2(r,t)}{1+2E(r)} \textrm{d}r^2+Y^2(r,t)(\textrm{d}\theta^2+\sin^2\theta \textrm{d}\phi^2) \,,
\end{equation}
where $Y(r,t)$ is the scale function, the prime denotes derivation with respect to the coordinate radius $r$ and the arbitrary function $E(r)$ is the curvature function which is related to the spatial Ricci scalar by ${\cal{R}} = -4(E Y)'/(Y^2Y')$~\cite{Marra:2008sy}. Throughout the paper we will loosely refer to $E$ when talking of spatial curvature in the LTB model.
The FLRW solution is recovered by setting $Y(r,t)\rightarrow a(t) \, r$ and $E(r)\rightarrow -k \, r^2/2$ throughout the equations.
Note that in the LTB space the transverse expansion rate $H_{T} \equiv \dot Y/Y$ will generally differ from the longitudinal expansion rate $H_{L} \equiv \dot Y'/Y'$.

The dynamics of the model is governed by the following equation \cite{Bondi:1947av}:
\begin{equation} \label{dynamics}
   \frac{\dot{Y}^2}{Y^2}=\frac{2 F(r)}{Y^3}+\frac{8\pi G}{3}\rho_\Lambda +\frac{2E(r)}{Y^2} \,,
\end{equation}
where the dot denotes derivation with respect to the coordinate time $t$ and $\rho_\Lambda = \Lambda / 8\pi G$ is the energy density associated with the cosmological constant. The arbitrary function $F(r)$ (actually a constant of integration) represents the effective gravitating mass and is related to the local dust energy density $\rho_M(r,t)$ through the equation $4\pi G \, \rho_M = F'/ (Y^2Y')$.
It is interesting to note that  in the last equation the curvature term is missing and so the gravitating mass -- which is the quantity that matters in matching the metric to the background -- differs from the invariant mass~\cite{Bondi:1947av}. This leads to ``strong'' backreaction~\cite{Kolb:2009rp} effects in LTB models as it implies that the expansion rate of a given shell (e.g., the border of the hole) is not sourced by the averaged mass within.
We will discuss this in \ref{curvatureLTB} for the case of Swiss-cheese models.

It is useful to rewrite Eq.~(\ref{dynamics}) in the following, more familiar form
\begin{equation} \label{dynamics2}
   H_T^2(r,t)=H_0^2(r)\left[\Omega_M(r)\left(\frac{Y_0(r)}{Y(r,t)}\right)^3+\Omega_\Lambda(r)+\Omega_K(r)\left(\frac{Y_0(r)}{Y(r,t)}\right)^2\right] ,
\end{equation}
where $H_0(r)\equiv H_T(r,t_0)$, $Y_0(r)\equiv Y(r,t_0)$ and the (present-day) density parameters are
\begin{eqnarray*}
  \Omega_M(r) & \equiv & {2 F(r) \over  H_{0}^{2}(r) Y_{0}^{3}(r)}  \,, \qquad 
  \Omega_\Lambda(r)  \equiv  {8\pi G \over  3} {\rho_{\Lambda} \over  H^2_0(r)}   \,, \label{omegamg}  \\
    \Omega_K(r) & \equiv & 1- \Omega_M(r) -   \Omega_\Lambda(r) =  {2E(r) \over H^2_0(r) Y^2_0(r)} \,.
\end{eqnarray*}
Eq.~(\ref{dynamics2}) can be used to determine the age of the universe at a radial coordinate~$r$:
\begin{equation} \label{age}
   t_0-t_{B}(r)=\frac{1}{H_0(r)}\int\limits^1_0 \frac{dx}{\sqrt{\Omega_M(r)x^{-1}+\Omega_K(r)+\Omega_{\Lambda}(r)x^2}}\;.
\end{equation}
One can constrain the models by requiring a simultaneous big bang, i.e., by setting $t_B(r)=0$.
Simultaneous big bang excludes decaying modes which would be strongly in contradiction with the inflationary paradigm \cite{Biswas:2006ub,Zibin:2008vj}.
Note, however, that an almost simultaneous big bang $t'_B(r)\approx 0$ could be used~\cite{Celerier:2009sv} as discussed in Section \ref{SNeo}.

LTB models feature three arbitrary functions. Within the present formalism they are taken as $\Omega_{M}(r)$, $t_{B}(r)$ and $Y_{0}(r)$.
One of these is but an expression of the gauge freedom, which we may fix by setting $Y_0(r) \propto r$  or, equivalently, $F(r) \propto r^{3}$.

\section{``Strong'' backreaction in exact Swiss-cheese models} \label{curvatureLTB}

We will now argue that ``strong'' backreaction~\cite{Kolb:2009rp} effects on the background evolution are small for realistic Swiss-cheese models based on LTB metrics which feature a void at the center. See Ref.~\cite{Sussman:2011na} for a discussion of more general LTB setups.

Let us start by calculating the average expansion rate of an LTB hole at a fixed time $t$: any deviation from the background value $H$ will signal strong backreaction effects due to the matter inhomogeneities.
This is formally done by averaging the expansion scalar $\theta=H_{L}+2 H_{T}$. It is easy to see that:
\begin{equation} \label{ave}
\langle \theta \rangle = {\int_{V} \theta \, dV \over V}=
\frac{\int_{0}^{r_{0}}dr \, \theta \, Y^{2}\, Y' / \sqrt{1+2 E}}
{\int_{0}^{r_{0}} dr \, Y^{2} \, Y' / \sqrt{1+2E}}
= {{\dot V}\over V} \stackrel{E \ll 1}{\simeq} 3 \, \left. {\dot Y \over Y} \right|_{r=r_{0}}
= 3  H  ,
\end{equation}
where $V$ is the proper 3-dimensional volume of the space slices, and we remind that $r_{0}$ and $l_{0}$ are comoving and proper radiuses of the hole, which is taken to be much smaller than the horizon: $ l_{0} \ll l_{\rm hor}$.
Eq.~(\ref{ave}) shows that if the curvature $E$ is small then the average expansion rate $\langle \theta \rangle /3 $ is close to the background value $H$ and the strong backreaction is negligible.
This is indeed the case of flat LTB models ($E(r)=0$) for which it is easy to show that the kinematical backreaction $Q_{D}$ of Ref.~\cite{Buchert:2007ik} is exactly zero no matter how the free functions are specified~\cite{Marra:2008sy}.

We have now to estimate the order of magnitude of $E$ for {\em realistic} LTB models with a void at the center.
Let us consider for simplicity an EdS background and fix the gauge by setting $Y_0(r) \propto r$.
It is useful to rewrite Eq.~(\ref{dynamics}) as:
\begin{equation} \label{nove}
E(r)= {1 \over 2} H_{T}^{2}(r,t) \, Y^{2}(r,t)  - \frac{F(r)}{Y(r,t)}  \,,
\end{equation}
which can be understood as that the total energy per unit of mass of the shell $r$ is given by its kinetic energy per unit of mass ($ H_{T}^{2} \, Y^{2} = \dot Y^{2}$) plus its potential energy per unit of mass due to the total gravitating mass up to the shell $r$.
Note that thanks to spherical symmetry one is able to define a potential energy also in cases far away from nearly Newtonian ones and that the potential energy is related to the curvature~\cite{Bondi:1947av}.

The curvature function $E$ vanishes at the center (generally it is $E \propto r^{2}$~\cite{Mustapha:1998jb}) and at the border where the hole is matched to the EdS metric, for which $E=0$ and so the two terms of Eq.~(\ref{nove}) cancel each other:
see, for example, Fig.~4 of Ref.~\cite{Marra:2007pm}.
Moreover, at the center of the hole there is a void and so the second term in Eq.~(\ref{nove}) is in magnitude $\lesssim$ as compared to the first one and so we can evaluate $E$ by considering the first term only.
Roughly, for realistic voids, it is $H_{T} \sim H$ and so we have:
\begin{equation} \label{eappro}
E \sim {1 \over 2} H^{2}  \,  l_{0}^{2}  \lesssim  \left( { l_{0} \over l_{\rm hor}} \right )^{2} \ll 1 \, ,
\end{equation}
that is, the curvature is indeed small for sub-horizon holes~\cite{Kolb:2009rp} (see also \cite{Mattsson:2010vq}).
In Eq.~(\ref{eappro}) we have evaluated $Y$ by taking its maximum value $Y(r_{0},t)= l_{0}$: we remind indeed that $Y=0$ at the origin and $Y'>0$ as $Y'=0$ would give shell crossing (we are assuming $F'(r)>0$).
We stress that even if the compensating overdense shell features matter contrasts $\gg 1$, still $F(r)$ is close to the EdS value as it is an integrated quantity which smoothly tends to the background value.
This shows how spherical symmetry severely restricts the effective inhomogeneities allowed in an LTB model.
As said earlier, the dynamics of the shell r depends only on the total mass within and does not depend on the mass outside, that is, density inhomogeneities are already averaged out as far as the dynamics of the shell is concerned (similar considerations apply to the curvature function $E$ which also is an average quantity as far as the spatial Ricci scalar is concerned).

Eq.~(\ref{eappro}) connects the curvature to the expansion rate in the void. It is possible to evade the latter constraint by considering LTB models which feature $H_{T} \gg H$.
These models, however, can only be matched to the background metric asymptotically and not at a finite radius $l_{0}$ as with Swiss-cheese models.
Such a strong expansion rate, besides being unrealistic, would indeed cause shell crossing in the overdense shell surrounding the void in a very short time interval.
As we have said before, the dynamics of the inner void shells is unaffected by the outer dense shells which will be (no matter how the free functions are adjusted) squeezed towards the border of the hole causing shell crossing in a time 
$\Delta t  \sim (l_{0}-l_{0}/2)/ (\Delta H \; l_{0}/2) \sim \Delta H^{-1} \sim H_{T}^{-1} \ll H^{-1}$, where we estimated the edge of the void to be at $l_{0}/2$ (and so its peculiar velocity to be $\Delta H \; l_{0}/2$) and the ``space'' to cover to be $l_{0}-l_{0}/2$.
See Ref.~\cite{Sussman:2010ew} for a discussion of the evolution of general LTB profiles.

We have shown, at least qualitatively, what causes the curvature to be small: $\Delta H \sim 1$ on sub-horizon scales $l_{0}$.
In other words, the curvature is small if so is the quantity $\Delta H \; l_{0}$, which in Ref.~\cite{Kolb:2009rp} was interpreted as a background peculiar velocity.
As said before, it is $\Phi \propto E$ \cite{Biswas:2007gi} and one could re-interpret the smallness of the strong backreaction as the fact that the inhomogeneous universe (in this case the Swiss-cheese model) can be described by means of a newtonian perturbed metric with a small potential $\Phi$.
The calculations of this Appendix, however, show that this is an assumption of the model and not a general feature of the inhomogeneous universe.
It is indeed clear that in the present case the curvature (and so the newtonian potential) is small {\em because} the (sub-horizon) inhomogeneities are matched to an {\em a priori} chosen background: in other words the strong backreaction is assumed small since the beginning by demanding small background peculiar velocities~\cite{Kolb:2009rp,Kolb:2008bn}.

\section*{References}

\end{document}